\documentclass[a4paper,11pt]{article}
\pdfoutput=1
\usepackage{jcappub}
\usepackage{bm}
\usepackage{soul}
\usepackage{latexsym}
\usepackage{dcolumn}
\usepackage{amsfonts,amssymb,amsmath}
\usepackage{graphicx,epsfig}
\usepackage{psfrag}
\usepackage{braket}
\usepackage{caption}
\usepackage{subcaption}
\usepackage{graphicx}
\usepackage{rotating}
\usepackage{hyperref}
\usepackage{tikz}
\usepackage{comment}
\usepackage{float}

\hypersetup{
	unicode=false,          
	pdftoolbar=true,        
	pdfmenubar=true,        
	pdffitwindow=false,     
	pdfstartview={FitH},    
	pdftitle={My title},    
	pdfauthor={Author},     
	pdfsubject={Subject},   
	pdfcreator={Creator},   
	pdfproducer={Producer}, 
	pdfkeywords={keyword1} {key2} {key3}, 
	pdfnewwindow=true,      
	colorlinks=true,       
	linkcolor=red,          
	citecolor=cyan,        
	filecolor=magenta,      
	urlcolor=green,           
	linktocpage=true
}

\newcommand{\edm}{{\rm EDM}} 
\newcommand{\cdm}{{\rm CDM}}

\newcommand{\dd}{\mathrm{d}} 
 
\newcommand{\colortextHEX}[2]{%
	\textcolor[HTML]{#1}{#2}%
}
\usepackage[framemethod=default]{mdframed}
\newmdenv[skipabove=7pt,
skipbelow=7pt,
rightline=false,
leftline=false,
topline=false,
bottomline=false,
backgroundcolor=gray!10,
linecolor=gray,
innerleftmargin=5pt,
innerrightmargin=5pt,
innertopmargin=5pt,
innerbottommargin=5pt,
leftmargin=0cm,
rightmargin=0cm,
linewidth=4pt]{eBox}

\makeatletter
\gdef\@fpheader{}
\makeatother

\begin{document}

\title{Exotic Dark Matter and the DESI Anomaly}

\author[1]{Matteo~Braglia,}
\author[2]{Xingang Chen,}
\author[2]{Abraham Loeb}

\affiliation[1]{Center for Cosmology and Particle Physics, New York University, 726 Broadway, New York, NY 10003, USA}
\affiliation[2]{Institute for Theory and Computation, Harvard-Smithsonian Center for Astrophysics, 60 Garden Street, Cambridge, MA 02138, USA}
\emailAdd{mb9289@nyu.edu}
\emailAdd{xingang.chen@cfa.harvard.edu}
\emailAdd{aloeb@cfa.harvard.edu}

\abstract
{Exotic dark matter (EDM) refers to a dark matter species whose equation of state deviates from zero at late times. This behavior enables it to model a variety of non-standard late-time cosmologies, offering alternatives to various dark energy (DE) models, especially when the DE sector violates the null energy condition. In this work, by fitting to CMB, BAO, and Supernovae (SNe) data 
and comparing models in a Bayesian approach, we show that simple models of exotic dark matter are statistically comparable to the $w_0w_a$CDM DE model in explaining the recent anomaly in the late-time cosmological evolution suggested by DESI and supernova observations, although in both classes of models the evidence against the $\Lambda$CDM model only appears when the DES-Y5 or Union3 SNe dataset is included.
The value of $H_0$ remains similar to that in the DE model, except in the no-SNe case, where the DE model predicts lower values than $\Lambda$CDM, thereby worsening the Hubble tension, whereas the EDM models yield values closer to that of $\Lambda$CDM, albeit with larger uncertainty. In addition, the EDM models predict a drastically different energy budget for the present-day universe compared to the standard model, and provide an explanation for a coincidence problem in the DE-model explanation of the DESI anomaly.
}
\maketitle
		
\setcounter{page}{1}		

\section{Introduction}

Recent data from DESI and supernova observations provide tantalizing evidence that the late-time evolution of the universe may deviate from the predictions of the standard cosmological model. Non-standard dark energy (DE) models are conventionally used to interpret such deviations
\cite{DESI:2025zgx,DES:2024jxu,DESI:2024mwx,DESI:2024aqx,DESI:2024kob,Ishak:2024jhs,DESI:2024hhd,DES:2025bxy,DESI:2025fii}.
However, {besides additional fine-tuning issues associated with DE models extending beyond the cosmological constant (such as maintaining the flatness of potentials)}, recent observations appear to favor a DE equation of state (EOS) which takes values $w_{\rm DE}<-1$ in the past and only recently crosses the so-called ``phantom" divide $w_{\rm DE}=-1$, implying a violation of the null energy condition \cite{Hawking:1973uf,Moghtaderi:2025cns}.

On the other hand, it has been shown in \cite{Chen:2025wwn} that the qualitative features of the anomaly observed by these experiments can also be explained if we keep the DE sector as a cosmological constant, but allow non-zero values of the dark matter EOS (taking values between $-1$ and 1) to emerge at late times. In addition, beyond this specific anomaly, it is shown that such dark matter models could also be used to model a variety of non-standard cosmologies, as alternatives to a range of DE models. 
Therefore, even if future data shift preferences regarding the features of the anomaly, these dark matter models would remain an important alternative class to distinguish from DE models.
This class of dark matter is referred to as exotic dark matter (EDM), and viable microscopic realizations can be constructed relatively easily \cite{Chen:2025wwn}.

Modifications to the standard model from EDM models vary widely depending on the details of the EOS and the values of other parameters. To fully explore their potential and assess their statistical significance, a full data comparison across the entire allowed parameter space is necessary. In this work, we compare some of the simplest EDM models with DESI, supernova data, {and a compressed CMB likelihood that captures the early-Universe CMB physics ~\cite{Lemos:2023xhs}}.
Since the EDM changes the late-time cosmological evolution, it is also interesting to examine its impact on the value of $H_0$, with particular attention to the implications for the Hubble tension problem \cite{DiValentino:2021izs,Freedman:2025nfr}.

Before outlining the structure of the paper, we wish to highlight an important point regarding our approach. Given the profound implications for fundamental physics suggested by the apparent deviation from $\Lambda$CDM indicated by DESI results, it is no surprise that considerable effort has been devoted in the literature to scrutinizing both the measurements themselves and their interpretation in terms of new physics \cite{Efstathiou:2024xcq, Efstathiou:2025tie,  Colgain:2024ksa, Mukherjee:2025ytj, Wang:2025mqz}. Such scrutiny is not only legitimate but also indicative of a healthy scientific debate.

Our approach, however, is intentionally more modest and focused. Since EDM is a new theoretical proposal, our goal here is simply to demonstrate that EDM can be easily misinterpreted as a conventional DE scenario. 
To this end, we adopt the simplified analysis framework used by the DESI collaboration in~\cite{DESI:2025zgx} for $\Lambda$CDM, DE, and EDM models, and perform a direct comparison of the results. 
{We emphasize that, within this framework, only the effects of EDM on the background expansion of the universe are compared with data. Other important issues, such as the impact of EDM on structure growth and its observational consequences, are left for future study.}

This paper is organized as follows. In Sec.~\ref{Sec:EDM models}, we present some of the simplest EDM models with two to four more parameters than the $\Lambda$CDM model. We first analyze these models analytically and numerically, to illustrate some of their most important effects on cosmology. This section also sets up notations used in subsequent data analyses, as well as provides some qualitative understanding of the data analysis results.
In Sec.~\ref{Sec:methodology}, we describe the datasets and the data analysis methods.
In Sec.~\ref{Sec:Results}, we compare all three models described in Sec.~\ref{Sec:EDM models} with CMB, BAO and supernova data. We present and describe the results of the data analyses, compare the statistical significances of different models using a Bayesian approach.
We conclude and discuss future directions in Sec.~\ref{Sec:Conclusions discussions}.

The following terminology is used throughout the paper. Models that include a component of exotic dark matter in addition to the standard $\Lambda$CDM components are referred to as $\Lambda$CEDM models, and in short EDM models.
The model that includes an evolving DE component parameterized by the Chevallier–Polarski–Linder (CPL) ansatz~\cite{Chevallier:2000qy,Linder:2002et} is called the $w_0w_a$CDM model, or in short the DE model.

\section{Models of exotic dark matter}
\label{Sec:EDM models}

\paragraph{A three (and two)-parameter model.} To understand the basic physics of EDM, let us consider a simple example introduced in Ref.~\cite{Chen:2025wwn} that is easily tractable analytically. Here we analyze this model in more detail to gain insights into how various parameters affect the cosmological evolution at late times and how they can produce effects similar to those of DE models. 

We consider a component of EDM of which a non-zero EOS emerges at late times, parameterized by a sharp step function at a transition redshift $z_t\equiv(1-a_t)/a_t$:
\begin{align}
    w_{\rm EDM} (a) = 
\begin{cases}
0 & \text{if } a < a_t ~,\\
w_{\edm,\,0} & \text{if } a\ge a_t ~.
\end{cases}
\label{Eq:EDMEOS}
\end{align}
We introduce another parameter $f_\edm$ to denote the fractional contribution of this EDM component to the total energy density of the combined CDM and EDM sector, {\em before} the emergence of the non-zero EOS:
\begin{align}
    f_\edm &\equiv 
        \lim_{a\to0} \frac{\rho_{\edm}(a)}{\rho_{\cdm}(a)+\rho_{\edm}(a)} ~.
    \label{Eq:fEDM_def}
\end{align}
The time coordinate is also often written in other variables, such as $N \equiv\ln a = -\ln(1+z)$, where $a$ denotes the scale factor and $z$ denotes the redshift. So this model has a total of three extra parameters relative to the standard $\Lambda$CDM model, namely, $f_\edm$, $a_t$, and $w_{\edm,\,0}$.

In data analyses, we will also consider a two-parameter limit of this model as $f_\edm\to 1$, and compare the results of this interesting limit in the context of more general cases.\footnote{\label{footnote_Giani}{Note that, for fitting the DESI DR2 + DES-Y5 data, the analysis of this two-parameter limit is similar to that of Ref.~\cite{Giani:2025hhs} (see also arXiv version 2 of Ref.~\cite{Giani:2024nnv}) which appeared recently. On the other hand, there are important differences. 
According to Ref.~\cite{Giani:2025hhs}, in their model the time-dependent modification to the background EOS is derived purely from back-reactions of halo or void formation, with the $\Lambda$CDM model still being the underlying theory. We would like to emphasize that, in a causal and local effective theory, the background evolution cannot be influenced by sub-horizon physics, e.g.~the timing of halo formation; otherwise, it would violate the principle of causality. Therefore, new physics beyond the $\Lambda$CDM model, such as various DE and EDM models, is required.}}

Our goal of this section is to gain a qualitative understanding on how the evolution of the Hubble parameter is modified by the presence of this component, relative to the evolution of the Hubble parameter in the $\Lambda$CDM model. This includes the effects of $f_\edm$, $a_t$, and $w_{\edm,\,0}$. As we will also be interested in the implication of EDM models on the Hubble tension problem, here, more general than \cite{Chen:2025wwn}, we allow the value of the present-day Hubble parameter in the EDM model to differ from that in the $\Lambda$CDM model. We denote these as $\tilde H_0$ and $H_0$, respectively. Note that we use a \~{} to distinguish parameters associated with non-$\Lambda$CDM models from their $\Lambda$CDM counterparts. But this notation will only be used in this subsection where confusion can arise. Accordingly, we will examine the impact of the ratio $\tilde H_0/H_0$ on the Hubble parameter evolution.

In the subsequent data analyses, we will fit both the $\Lambda$CDM model and an alternative model, such as the EDM or DE model, to the same sets of data.
Here, we would like to analytically understand why an EDM model would fit better or worse than the $\Lambda$CDM model, as well as the range of the viable parameter space of this model, by comparing the difference of the model predictions with the residual behavior of the data relative to the $\Lambda$CDM model. 

For the $\Lambda$CDM model, We will use a fixed set of best-fit parameters, namely the present-day density parameters ($\Omega_b=0.049,\Omega_{\cdm}=0.271,\Omega_\Lambda=0.68$), mainly constrained by the CMB data. 
For the EDM model, setting the sound speed of EDM to zero (see \cite{Chen:2025wwn} for a discussion), EDM behaves just as CDM before the emergence of the non-zero EOS value. The early-time constraints on the EDM model again mainly come from CMB anisotropies, and they should be essentially the same as those on the $\Lambda$CDM model. While a realistic data analysis may lead to small differences, for the purpose of this section it is sufficient to choose the parameters of the EDM model such that they reproduce the early-time evolution of the best-fit $\Lambda$CDM model.\footnote{Several recent studies have explored scenarios in which dark matter takes other forms of non-zero equation of state, motivated by the latest DESI results. In particular, Ref.~\cite{Wang:2025zri} considered a CPL-like parameterization for cold dark matter, while Refs.~\cite{Kumar:2025etf,Li:2025eqh,Li:2025dwz} investigated models with a constant shift in the DM equation of state. Compared to EDM models, the values of dark matter EOS in these models are not turned on only at late-times, thus are much more tightly constrained due to early Universe CMB physics. {Therefore, these models are expected to fit the DESI anomaly worse than the $w_0w_a$CDM model.} }

Denoting the present-day density parameters of the $\Lambda$CDM model as  $\Omega_\cdm$, $\Omega_b$, and $\Omega_\Lambda$, and the critical density as $\rho_{c,0}$, at $a=a_t$ the density of these components should be
\begin{align}
    \rho_{\cdm,\,t} &= \Omega_\cdm \rho_c a_t^{-3} ~,\\
    \rho_{b,\,t} &= \Omega_b \rho_c a_t^{-3} ~,\\
    \rho_{\Lambda,\,t} &= \Omega_\Lambda \rho_c ~.
\end{align}
Therefore, at $a=a_t$, the densities of various components of the matching EDM model, with the EOS \eqref{Eq:EDMEOS} and the EDM fraction parameter $f_\edm$ \eqref{Eq:fEDM_def}, should be
\begin{align}
    \tilde \rho_{\cdm,\,t} &= (1-f_\edm) \rho_{\cdm,\,t} 
    = (1-f_\edm) \Omega_\cdm \rho_c a_t^{-3} ~,\\
    \tilde \rho_{\edm,\,t} &= f_\edm \rho_{\cdm,\,t} 
    = f_\edm \Omega_\cdm \rho_c a_t^{-3} ~,\\
    \tilde \rho_{b,\,t} &= \rho_{b,\,t} = \Omega_b \rho_c a_t^{-3} ~.
    \label{Eq:densities at a1}
\end{align}
On the other hand, the density of the cosmological constant, $\tilde\rho_{\Lambda,\,t}$, is a free parameter, because its value is negligible during the recombination.

Evolving these values to present-day, $a=a_0=1$, we have
\begin{align}
    \tilde \rho_{\cdm,0} &= (1-f_\edm) \Omega_\cdm \rho_c ~,\\
    \tilde \rho_{\edm,0} &= f_\edm \Omega_\cdm \rho_c a_t^{3w_{\edm,\,0}} ~,\\
    \tilde \rho_{\rm B,0} &= \Omega_b \rho_c ~,
    \label{Eq:densities at a0}
\end{align}
and we also define $\tilde\rho_{\Lambda,0} \equiv \beta_\Lambda \rho_c$.

So, the present-day critical density of this matching EDM model is
\begin{align}
    \tilde\rho_c = \left[ \beta_\Lambda + \Omega_b + (1-f_\edm)\Omega_\cdm
    + f_\edm a_t^{3 w_{\edm,\,0}} \Omega_\cdm \right] \rho_c ~,
\end{align}
and the present-day density parameters of this model are
\begin{align}
    \tilde\Omega_\cdm &= (1-f_\edm) \Omega_\cdm \alpha^{-1} ~, \\
    \tilde\Omega_{\edm} &= f_\edm a_t^{3w_{\edm,\,0}} \Omega_\cdm \alpha^{-1} ~, \\
    \tilde\Omega_b &= \Omega_b \alpha^{-1} ~, \\
    \tilde\Omega_\Lambda &= \beta_\Lambda \alpha^{-1} ~,
\end{align}
where we have defined
\begin{align}
    \alpha\equiv \frac{\tilde\rho_c}{\rho_c} = 
    \frac{\tilde H_0^2}{H_0^2}
    =
    \beta_\Lambda + \Omega_b + (1-f_\edm) \Omega_\cdm 
    + f_\edm a_t^{3w_{\edm,\,0}} \Omega_\cdm ~.
    \label{Eq:alpha_definition}
\end{align}

\begin{figure}[t]
		\includegraphics[width=\columnwidth]{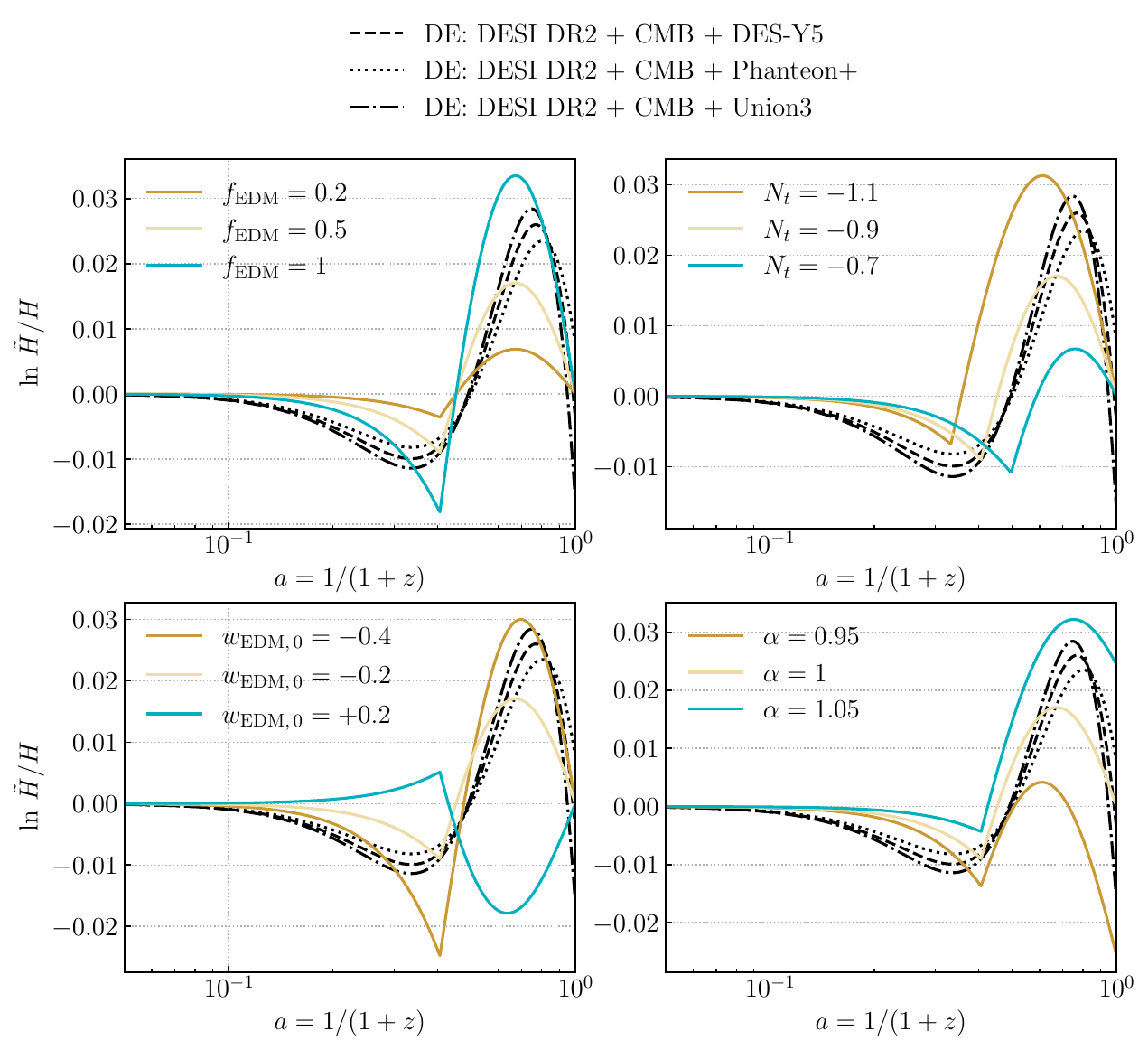}
  \caption{{\em Effects of $f_\edm,\, N_t,\, w_{\edm,\,0}$, and $\alpha\equiv \tilde H_0^2/H_0^2$ on Hubble parameter evolution.}
   Each colored line shows the log-ratio of the Hubble parameters between an EDM model and a best-fit $\Lambda$CDM model, $\ln(\tilde H/H)$.
   The parameters of the $\Lambda$CDM model is fixed in all cases ($\Omega_b=0.049,\Omega_{\cdm}=0.271,\Omega_\Lambda=0.68$).
   {
   In all four panels, the {\bf \colortextHEX{EDDCA5}{light gold} }curve represents the model with $f_\edm=0.5, N_t=-0.9, w_{\edm,\,0}=-0.2,\alpha=1$, and other density parameters are chosen to match the early-time behavior of the $\Lambda$CDM model (see main text for method). }
   In the top-left panel, we vary the fraction of EDM in DM at early-time, $f_\edm=0.2,0.5,1$. In the top-right panel, we vary the timing of transition, $N_t=-1.1,-0.9,-0.7$. In the bottom-left panel, we vary the magnitude of EOS step, $w_{\edm,\,0}=-0.4,-0.2,+0.2$. In the bottom-right panel, we vary the Hubble parameter ratio, $\alpha=0.95,1,1.05$.
   The black dashed, dotted and dash-dotted represent the mean $w_0w_a$CDM models for the DESI DR2 + DES-Y5, Pantheon+, or Union3 data respectively, see~\cite{DESI:2025zgx} for parameter values.
  }
  \label{fig:3-para-models}
\end{figure}

Therefore, fixing the parameters of the $\Lambda$CDM model ($\Omega_\Lambda$, $\Omega_\cdm$, $\Omega_b$), and given the values of the EDM model parameters ($f_\edm$, $a_t$, $w_{\edm,\,0}$) along with $\alpha$, we can determine the present-day density parameters of the EDM model, use them to compute its Hubble parameter evolution through the following Friedmann equation,
\begin{align}
    \frac{\tilde H^2}{\tilde H_0^2} = 
    \begin{cases}
    (\tilde\Omega_b+ \tilde\Omega_\cdm) a^{-3} + \tilde\Omega_{\rm EDM}~ a_t^{-3w_{\edm,\,0}} a^{-3} + \tilde\Omega_\Lambda 
    & \text{if } a < a_t ~,\\
    (\tilde\Omega_b+ \tilde\Omega_\cdm) a^{-3} + \tilde\Omega_{\rm EDM}~ a^{-3(1+w_{\edm,\,0})} + \tilde\Omega_\Lambda 
    & \text{if } a\ge a_t ~,
    \end{cases}
    \label{eq:Friedmann_EDM models}
\end{align}
take a logarithm and subtract the $\Lambda$CDM evolution from it. The above procedure guarantees that the early-time difference will vanish as $a\to0$, and what's left is their late-time difference including that of the present-day Hubble parameter values. In Fig.~\ref{fig:3-para-models}, we give several examples of the effects of each parameter on the Hubble parameter evolution.

We also would like to qualitatively compare these examples with the preference from data. In this section, for simplicity, we plot several DESI DR2's best-fit models of the DE models \cite{DESI:2025zgx} as proxies of the data preference. The residual Hubble evolution relative to the $\Lambda$CDM model shown in these examples is also supported by results using a binning reconstruction approach \cite{DESI:2025zgx}. To plot the difference between the DE models and $\Lambda$CDM models, we again use the condition that they have to match at early times. In this case, this condition is simply 
$\tilde\Omega_{m} \tilde\rho_c = \Omega_{m} \rho_c$, which means
\begin{align}
    \frac{\Omega_{m}}{\tilde\Omega_{m}} = \frac{\tilde H_0^2}{H_0^2} ~.
\end{align}
We can use this relation to ensure that the early-time difference vanishes, e.g., by determining $\Omega_{m}$ from $\tilde\Omega_{m}$, $\tilde H_0$, and $H_0$. The evolution of $\tilde H(a)$ is determined by the following Friedmann equation with the DE taking the CPL ansatz~\cite{Chevallier:2000qy,Linder:2002et}:
\begin{align}
    \frac{\tilde H^2}{\tilde H_0^2} = 
    \tilde\Omega_{m} a^{-3} + \tilde\Omega_\Lambda e^{-3w_a(1-a)}
    a^{-3(1+w_0+w_a)} ~.
\end{align}
Three best-fit DE models are plotted in Fig.~\ref{fig:3-para-models}, to compare with the EDM models. The comparison helps to understand how EDM models could mimic the effects of the DE models and fit the same data.

Besides the obvious dependence on several parameters illustrated in Fig.~\ref{fig:3-para-models}, the following several observations are worth noting and are helpful to understand some of the results in the subsequent data analyses.
\begin{itemize}
    \item As shown in \cite{Chen:2025wwn}, the primary consequence of a negative $w_{\edm,\,0}$ is a bump in $\ln(\tilde H/H)$. On the other hand, in most examples, there is also a shallow trench before $a$ reaches $a_t$. This feature is due to the freedom in assigning the value of the cosmological constant mentioned below \eqref{Eq:densities at a1} and \eqref{Eq:densities at a0}, and parameterized by $\beta_\Lambda$. 
    In most examples in Fig.~\ref{fig:3-para-models}, $\tilde\Omega_\Lambda$ of the EDM model is lower than $\Omega_\Lambda$ of the $\Lambda$CDM model; this difference is negligible at early times but becomes noticeable as the universe approaches later times, creating the trench before reaching $a=a_t$. In this model, it turns out that both the trench and bump receive interesting supports in data fitting -- the former mimics the DE model in part of the $w<-1$ region and the latter mimics the phantom crossing and the $w>-1$ part of the DE model. 

    \item Comparing the top-left to the bottom-left panel, we can see that there is a degeneracy in the effects of $f_\edm$ and $w_{\edm,\,0}$. Namely, increasing $f_\edm$ or increasing $|w_{\edm,\,0}|$ ($w_{\edm,\,0}<0$) share some similarities in their effects on $\tilde H^2/H^2$.
    
    \item The bottom-right panel demonstrates some trend on how this EDM model could impact the value of $\tilde H_0$. The freedom in $\beta_\Lambda$ introduces some flexibility in the present-day Hubble parameter, $\tilde H_0$, relative to the $\Lambda$CDM value, as indicated by Eq.~\eqref{Eq:alpha_definition}. However, this freedom will be correlated with other features in the evolution mentioned previously. Overall, although this model could lead to a range of $\tilde H_0$ values, the height of the bump allowed by the data restricts how large $\tilde H_0$ can reach and the current error bars on the late-time cosmology also increase its uncertainties. Our subsequent data analysis will provide a more precise determination of its value.
\end{itemize}

\begin{figure*}[t!]
	\begin{center}
		\includegraphics[width=\columnwidth]{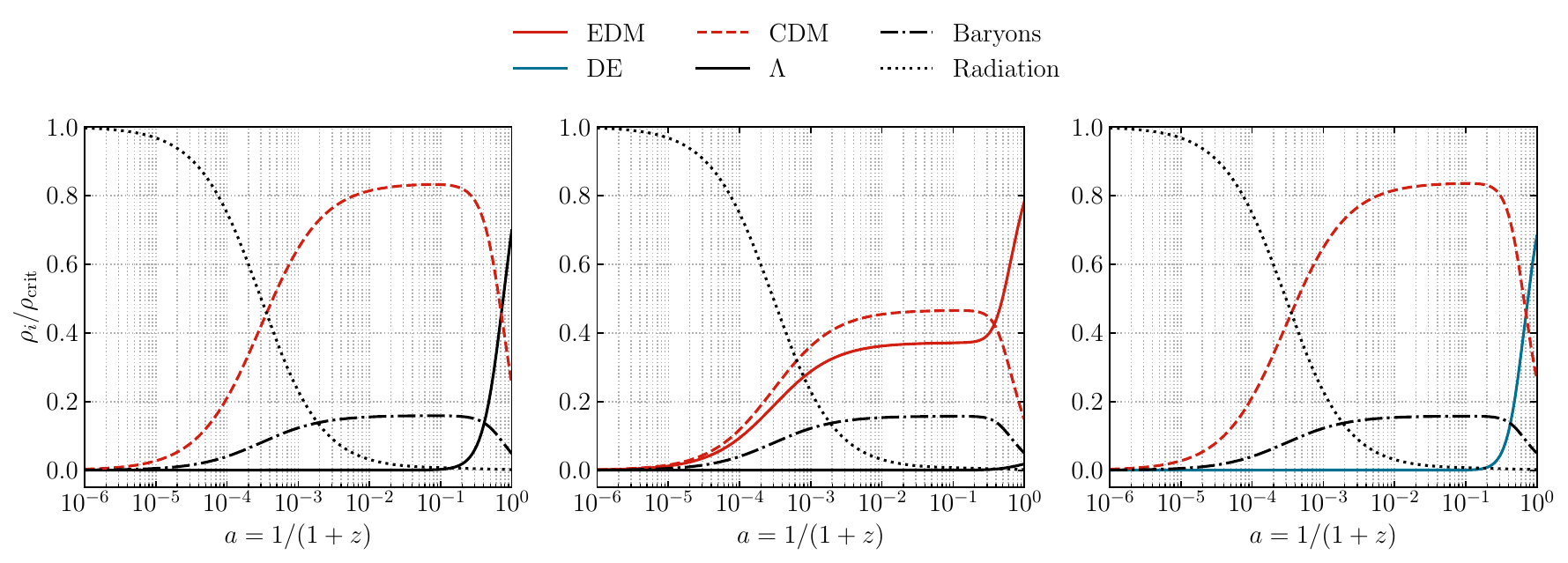}
	\end{center}\caption{\footnotesize\label{fig:intro} Evolution of the energy fractions in the $\Lambda$CDM model [left], the EDM model [middle], and the DE model [right]. We use the best-fit parameters from the  subsequent analysis of  the DESI DR2 + CMB + DES-Y5 data set, see Table~\ref{tab:constraints_D5} in Sec.~\ref{Sec:Results}.}
    \label{Fig:components_evolution}
\end{figure*}

It is also informative to compare the evolution of various components in the EDM model with those in the $\Lambda$CDM and DE (i.e.~$w_0w_a$CDM) models, respectively. Figure~\ref{Fig:components_evolution} illustrates such a comparison between  the $\Lambda$CDM model, the EDM model, and the DE model. Between the $\Lambda$CDM and DE model, the differences in their late-time evolution only cause small differences in the predicted energy densities. However, for the EDM model, the differences lead to striking changes in the energy densities. For instance, while providing similarly good fits to current datasets, a best-fit EDM model predicts a present-day DE density parameter of only about 6\% of the total energy density -- significantly lower than in either the DE or $\Lambda$CDM models. Moreover, in the EDM scenario, equality between DE and dark matter (CDM+EDM) has not yet occurred.
While this behavior may appear specific to the particular parameter choice shown in Fig.~\ref{Fig:components_evolution}, we will show that it is a general feature shared by all best-fit solutions of the EDM models considered in this paper.

This contrast also has conceptual implications. In the DE model, current data lead to an extra coincidence problem: the phantom crossing and the DE-dark matter equality occur around the same time. In the EDM model, however, this coincidence vanishes -- there is only a single process, the evolution of EDM's EOS, which is reinterpreted as two separate but simultaneous phenomena in the DE framework.

\paragraph{A four-parameter model.} We now introduce one additional parameter to smooth out the step-function in \eqref{Eq:EDMEOS}. This defines the most general model considered in this paper. We also establish notation that will be more conveniently used in the subsequent data analysis.

We modify \eqref{Eq:EDMEOS} to the following form:
\begin{equation}
	\label{eq:wEDM}
	w_\edm(N) =   \frac{w_{\edm,\,0}}{2}\left[1-\tanh\frac{N_t-N}{\sigma_{\edm}}\right],
\end{equation}
where again $N\equiv \ln a$. The parameter $\sigma_{\edm}$ controls the sharpness of the transition of the EDM's EOS from $0$ (before $z_t$) to $w_{\edm,\,0}$ (after $z_t$). By taking appropriate limits, this four-parameter model can be easily reduced to the previous two-, or three-, parameter versions. In our data analyses we will consider all three cases.

The time evolution of the EDM energy density is easily found by solving the continuity equation, and reads:
\begin{eqnarray}
	\rho_\edm(a) = \rho_{\edm,\,0}\exp\left[ 3\int^{\ln a_0\equiv 0}_{\ln a}\dd N' \left(1+w_\edm(N')\right) \right]\equiv \rho_{\edm,\,0}\left(\frac{a_0}{a}\right)^3 e^{\mathcal{I}(a)},
\end{eqnarray} 
where the subscript ``${}_0$" in $\rho_{\edm,\,0}$ means its value evaluated at present-day $a=a_0=1$, or equivalently $z=0$. In the equation above, we have defined
\begin{align}
	\mathcal{I}(a)&=\int^{\ln a_0\equiv0}_{\ln a}\dd N'\,\frac{3 w_{\edm,\,0}}{2}\left[1-\tanh\frac{N_t-N'}{\sigma_{\edm}}\right]\notag\\
	&=\frac{3w_{\edm,\,0}}{2}\left\{-N-\sigma_{\edm}\ln\frac{\cosh\frac{N-N_t}{\sigma_{\edm}}}{\cosh\frac{N_t}{\sigma_{\edm}}}\right\},
\end{align}
or
\begin{equation}
	e^{\mathcal{I}(a)} = a^{-\frac{3w_{\edm,0}}{2}} \left[ \frac{a_t^{-\frac{1}{\sigma_{\edm}}} + a_t^{\frac{1}{\sigma_{\edm}}}}{ \left( \frac{a}{a_t} \right)^{\frac{1}{\sigma_{\edm}}} + \left( \frac{a}{a_t} \right)^{-\frac{1}{\sigma_{\edm}}} } \right]^{\frac{3w_{\edm,0}\sigma_{\edm}}{2}} ~.
\end{equation}
As expected, the factor $e^{\mathcal{I}(a)}$ approaches a constant before the transition in the EOS, which confirms that the EDM density evolves as a matter fluid, and approaches $a^{-3w_{\edm, 0}}$ after the transition, so that the fluid redshifts faster or slower than matter depending on whether $w_{\edm, 0}$ is positive or negative respectively.

Let us also make some further comments on the abundance of the EDM. EDM has important effects on the background expansion both before and after the {transition} redshift $z_t$. Due to the change in its EOS, the abundance of EDM evolves with time differently from CDM. In the following, we relate the parameters that describe its abundance in early- and late-time limit respectively.

As in the previous case, we use $f_\edm$ to denote the fraction of the EDM density, $\rho_{\edm}$, in all dark matter density, $\rho_{\cdm}+\rho_{\edm}$, in the early-time limit. To capture the contribution of EDM to the early cosmic history, this parameter relates them at $z\gg z_t$, when their ratio is constant as they both redshift as $\propto a^{-3}$. By construction, the parameter is bounded by $0\leq f_\edm\leq1$ and given by:

\begin{equation}
        f_\edm \equiv 
        \lim_{a\to0} \frac{\rho_{\edm}(a)}{\rho_{\cdm}(a)+\rho_{\edm}(a)} \equiv\frac{r_\edm}{1+r_\edm} ~,
        \label{eq:fEDM_def}
\end{equation}
where we have defined the early-time EDM to CDM ratio as
	\begin{align}
	r_\edm&\equiv\lim_{a\to0}\frac{\rho_\edm(a)}{\rho_\cdm(a)}=\lim_{a\to0}\frac{\rho_{\edm,\,0}}{\rho_{\cdm,\,0}}e^{\mathcal{I}(a)}\notag\\
        &=\frac{\Omega_{\edm}}{\Omega_\cdm} e^{-\frac{3}{2} w_{\edm,\,0} N_t}
          \left( 2 \cosh\frac{N_t}{\sigma_{\edm}} \right)^{\frac{3}{2}w_{\edm,\,0}\sigma_{\edm}}\equiv \frac{\Omega_{\edm}}{\Omega_\cdm}e^{\mathcal{I}_0} ~,
          \label{eq:rEDM_def}
\end{align}
where, as defined previously, $\Omega_i\equiv\rho_{i,\,0}/\rho_{{\rm crit},\,0}$.
We stress again that both $f_\edm$ and $r_\edm$ are defined at early time when the EOS of EDM is zero. Eq.~\eqref{eq:rEDM_def} relates $r_\edm$ to the conventional definitions of density parameters, $\Omega_\cdm$ and $\Omega_{\edm}$, which are defined with density values today. The parameter we have just introduced is crucial to understand the effects of EDM on the physics around the time of recombination, responsible for the production of the CMB anisotropies. As well-known, the standard-model early universe physics of the anisotropies is mainly controlled by 3 parameters: the angular  scale of the acoustic
fluctuations, $\theta_*=D_M(z_*)/r_*$, where $r_*$ is the comoving
sound horizon at recombination, and the (reduced) baryon and CDM energy fractions, $\omega_b$ and $\omega_c$ (see e.g.~\cite{Lemos:2023xhs}). Eqs.~\eqref{eq:fEDM_def} and~\eqref{eq:rEDM_def} show that, in the EDM model, the {\em effective} CDM energy fraction that is experienced by the CMB at early times is 
	\begin{equation}
    \label{eq:omega_eff_early}
		\omega^{\rm eff}_{c}\equiv\omega_c(1+r_\edm).
	\end{equation}

On the other hand, we could also interpret Eq.~\eqref{eq:rEDM_def} as a relation between the present day density parameter of EDM and that of CDM, given the parameters $w_{\edm,\,0},\,\sigma_{\rm EDM},\,z_t$, which is a way to measure the impact of EDM on the late time expansion:
	\begin{equation}
		\Omega_{\edm}= \Omega_{\cdm}\,\frac{f_{\edm}}{1-f_{\edm}}\,e^{-\mathcal{I}_0}\left(w_{\edm,\,0},\,\sigma_{\rm EDM},\,z_t\right). 
	\end{equation}

The model is therefore described by 4 parameters: $\Omega_{\edm}$ (or $f_\edm$), $\,w_{\edm,\,0},\,\sigma_{\rm EDM},\,z_t$. These parameters describe the background evolution of the effective fluid. In our analysis, we will be using $\Omega_{\edm}$ instead of $f_\edm$, the latter of which is not a convenient parameter to sample over. The reason is that, upon fixing the factor $f_\edm$, $e^{-\mathcal{I}_0}$ can be arbitrarily large, easily making the sampler explore unphysical regions where $\Omega_{\rm EDM}>1$. On the other hand, one can choose a very simple and physically motivated prior and compute $f_\edm$ as a derived parameter using Eqs.~\eqref{eq:fEDM_def} and~\eqref{eq:rEDM_def}. In practice, since $\Omega_{\cdm}$ can also be quite large, we will employ the parameter $\tilde{\Omega}_{m}\equiv\Omega_{b}+\Omega_{\cdm}+\Omega_{\edm}$. In addition to these background parameters, the EDM perturbations are characterized by its speed of sound $c_s^2$. In \cite{Chen:2025wwn}, it is shown that $c_s$ of EDM can be easily made small. So for this paper, we simply assume $c_s^2=0$ and defer an exploration of the effects of the speed of sound to future works. 

To compare the effects of EDM to those of DE, in this paper we will again adopt the CPL parameterization~\cite{Chevallier:2000qy,Linder:2002et} in the DE model:
\begin{equation}
	w_{\rm DE}(a)= w_0 + w_a (1-a). 
\end{equation}
The CPL parameterization is essentially a first-order Taylor expansion of the DE EOS around the present time, $a_0=1$. While it has been routinely used in the literature to constrain DE with cosmological data (see~\cite{Cortes:2024lgw,Colgain:2024xqj,Shlivko:2024llw,Park:2024vrw,Giare:2024gpk,Giare:2024oil,RoyChoudhury:2024wri, Colgain:2024mtg,Giare:2025pzu, RoyChoudhury:2025dhe} for an incomplete list of recent papers), following DESI DR1~\cite{DES:2024jxu} and early hints of a dynamical DE component, the efficiency of the CPL form for data analysis has been debated. Some studies argue that it accurately captures the phenomenology of various physically motivated models, such as quintessence~\cite{Tada:2024znt,Wolf:2024eph,Wolf:2025jlc,Bayat:2025xfr} and models with non-minimal couplings~\cite{Rossi:2019lgt,Ferrari:2023qnh,Wolf:2024stt,Pan:2025psn,Wolf:2025jed,Ferrari:2025egk}. Others contend that certain theoretical models~\cite{Bhattacharya:2024hep,Ramadan:2024kmn,Notari:2024rti,Bhattacharya:2024kxp,Borghetto:2025jrk,Payeur:2024dnq,Dinda:2025iaq,Gialamas:2025pwv,Cline:2025sbt} and features in current data cannot be fully captured by such a simple expansion~\cite{Dinda:2024ktd,Ye:2024ywg,Nesseris:2025lke,Ormondroyd:2025exu,Ormondroyd:2025iaf,Berti:2025phi,Gonzalez-Fuentes:2025lei}. Our goal here is not to weigh in on that debate, but to demonstrate that the phenomenological EDM model defined in Eq.~\eqref{Eq:EDMEOS} — though also simplified — can reproduce features similar to those arising from the CPL parameterization.

\section{Data analysis: methodology and data description}
\label{Sec:methodology}

In this section, we describe the methodology used in our data analysis. We employ a modified version of the \texttt{CLASS} code~\cite{Lesgourgues:2011re,Blas:2011rf}, extended to include the evolution of the EDM fluid as described by Eq.~\eqref{eq:wEDM}. For the $\Lambda$CDM baseline, we sample the standard cosmological parameters $\{H_0,\,\omega_b,\,\omega_c\}$. In the EDM and DE models, we additionally sample over $\{w_{\edm,\,0},\,\tilde{\Omega}_m,\,z_t,\,\log_{10}\sigma_\edm\}$ and $\{w_0,\,w_0 + w_a\}$, respectively.


We note that in the EDM model, the EDM component contributes to CDM at early times, so $\omega_c^{\rm eff} = \omega_c (1+r_\edm)$ (see Eq.~\eqref{eq:omega_eff_early}), whereas in $\Lambda$CDM and DE, $\omega_c^{\rm eff} = \omega_c$ trivially. For the extra parameters, we assume the following flat priors:

\begin{table}[H]
\centering
\begin{tabular}{l c}
\hline
Parameter & Prior range \\
\hline
$w_{\edm,\,0}$ & $[-1,\,1]$ \\
$\tilde{\Omega}_m$ & $[0,\,1]$ \\
$z_t$ & $[0,\,10]$ \\
$\log_{10}\sigma_{\edm}$ & $[-2,\,0.5]$ \\
$w_0$ & $[-3,\,1]$ \\
$w_0 + w_a$ & $[-6,\,0]$ \\
\hline
\end{tabular}
\end{table}

In particular, the prior on the EOS is justified as follows: the lower limit is set by the theoretical requirement of not crossing the phantom divide, whereas we have tested that for $w_{\edm,\,0}>1$ the EDM fluid redshifts away so quickly that its effects are phenomenologically indistinguishable from $w_{\edm,\,0}=1$.

Since the parameter spaces of the extended models are nontrivial and no detection of additional parameters is guaranteed a priori, the resulting posterior distributions are typically non-Gaussian and potentially multimodal. To accurately explore such complex structures, we use nested sampling as implemented in \texttt{polychord}~\cite{Handley:2015fda,Handley:2015vkr}. In order to allow an efficient exploration of the parameter space we use 1000 live points in our analysis, corresponding to $\sim167$ points per dimension of the parameter space for the model with highest number of parameters, i.e. EDM 4p.  In addition to providing parameter constraints, \texttt{polychord} computes the Bayesian evidence $\ln \mathcal{Z}_i$ for each model $i$, enabling model comparison via the Bayes factor $\ln B_i \equiv \ln \mathcal{Z}_i - \ln \mathcal{Z}_{\Lambda{\rm CDM}}$. 

Finally, after performing the Bayesian analysis, we identify the best-fit parameters using the \texttt{bobyqa} minimizer~\cite{powell2009bobyqa}. {For each of the best-fit parameters of the extended models, we report the corresponding $\Delta\chi^2=\chi^2-\chi^2_{\Lambda {\rm CDM}}$, as well as the so-called Akaike information criterion (AIC) defined as $\Delta{\rm AIC}\equiv\Delta\chi^2+2\Delta p$, where $\Delta p$ is the number of extra parameters~\cite{Liddle:2007fy}.} Our full analysis pipeline, including the likelihoods for the datasets described below, is implemented in the \texttt{COBAYA} code~\cite{Torrado:2020dgo}. {Finally, we use the samples from the nested sampling to plot the so-called predictive posterior distribution of the EOS for both DE and EDM models analysis, which translates the constraints on the model parameters into the constraints on the EOS as a function of redshift. Our plots are produced using the code fgivenx~\cite{fgivenx}.}

\paragraph{DESI DR2 BAO:} We use BAO measurements from the second data release (DR2) of DESI~\cite{DESI:2025qqy,DESI:2025zgx}. These include determinations of the Hubble distance $D_H(z)/r_d$ and the comoving angular diameter distance $D_M(z)/r_d$ over the redshift range $0.51 < z < 2.33$. In addition, we include the measurement of the angle-averaged BAO scale,
$$
D_V/r_d = \left( z_{\rm eff} D_M^2 D_H \right)^{1/3} / r_d,
$$
evaluated at the effective redshift $z_{\rm eff} = 0.295$.

\paragraph{DES-Y5 SNe:} We use the DES-Y5 sample, which includes 1,635 light curves from 1,550 Type Ia supernovae (SNe Ia) in the range $0.10 < z < 1.13$, as well as 194 low-redshift SNe Ia in the range $0.025 < z < 0.10$. This dataset constrains the expansion history through measurements of the luminosity distance $D_L(z)$. We note that most of the evidence in favor of DE arises from the inclusion of low-redshift data~\cite{Notari:2024zmi,DESI:2025zgx,DESI:2025fii,Huang:2025som}. However, since the goal of this paper is to show that the EDM model can mimic DE, we do not perform an detalied analysis on partial subsets of this data and instead use the full sample throughout.

\paragraph{Pantheon+ SNe:} As an alternative supernova dataset, we also use the Pantheon+ sample~\cite{Scolnic:2021amr}, which consists of 1,550 SNe Ia spanning the redshift range $0.001 < z < 2.26$.

\paragraph{Union3 SNe:} Our third and last alternative supernova dataset is the Union3 sample~\cite{Rubin:2023jdq}, which consists of 2,087 SNe Ia spanning the redshift range $0.05 < z < 2.26$.

\paragraph{CMB compressed.} 
The datasets discussed so far constrain only the late‑time expansion history of the Universe.  
To ensure that our models remain viable at early times, we incorporate information from the CMB.  
We follow the DESI analysis of Ref.~\cite{DESI:2025zgx}: instead of using the full CMB likelihood, we adopt a compressed approach in which the CMB data are distilled into a multivariate Gaussian prior on the early‑Universe parameters $\theta_s,\;\omega_b,\;\omega_c$.  

\medskip
\noindent
\begin{equation}
\boldsymbol{\mu}\bigl(100\,\theta_s,\;\omega_b,\;\omega_c^{\rm eff}\bigr)=
\begin{pmatrix}
1.04103\\[2pt]
0.02223\\[2pt]
0.1192
\end{pmatrix},
\end{equation}
\begin{equation}
	\label{eq:prior_thetas}
	\boldsymbol{\Sigma}=10^{-8}\times
	\begin{pmatrix}
		6.62099420 &  1.24442058 & -13.1731741\\
		1.24442058 &  2.13441666 & -11.5345007\\
		-13.1731741 & -11.5345007  &  169.776300
	\end{pmatrix}.
\end{equation}

\noindent
These early-Universe quantities can be determined independently of late‑time assumptions by marginalizing over integrated Sachs–Wolfe (ISW) and CMB‑lensing effects.  
The covariance above is taken from Ref.~\cite{Lemos:2023xhs}, which derives it from the \texttt{CamSpec} likelihood and thereby greatly accelerates our analysis\footnote{We note that this differs slightly from the prior choice in Ref.~\cite{DESI:2025zgx}, which uses $\theta_*,\;\omega_b,\;\omega_{bc}=\omega_b+\omega_c$, along with  a different mean and covariance (see their Eqs.~(A.1) and (A.2)).  
With the likelihood adopted here, our results agree precisely with those of Ref.~\cite{DESI:2025zgx} (see Tables~\ref{tab:constraints_DR2}–\ref{tab:constraints_P+}) for the $\Lambda$CDM and DE models. The very small differences between our results and those in Ref.~\cite{DESI:2025zgx} likely arise from the slightly different definitions of $\theta_s$ and $\theta_*$ in \texttt{CLASS} and \texttt{CAMB}.}. 

In the following, we consider four combinations of the datasets described above. Our baseline is the DESI DR2 + CMB combination, which allows us to investigate the implications of the recent DESI BAO measurements while anchoring the fit to early-Universe information from the CMB. We then explore the impact of supernova data by separately adding our three SNe samples to the baseline.

To better understand the impact of each EDM parameter in the data analysis, we benchmark three different scenarios of increasing complexity --- i.e., with an increasing number of free parameters. The simplest case, denoted EDM 2p, introduces two additional parameters, just like the DE model. In this setup, we fix $f_\edm = 1$, effectively removing the CDM component\footnote{In practice, we set $\Omega_c = 10^{-4}$ for numerical stability.}, and fix the transition to be very sharp, $\log_{10}\sigma_\edm = -{2}${, i.e.~to the lower end of the prior of $\sigma_\edm$ in the 4p model}. The model is then described by the parameters ${\tilde{\Omega}_m,\, w_{\edm,\,0},\, z_t}$. The $\Lambda$CDM limit of this model corresponds to $w_{\edm,\,0} = 0$, in which case $z_t$ becomes irrelevant and $\tilde{\Omega}_m$ reduces to the standard $\Omega_m$.
Next, we consider a more general scenario, EDM 3p, where we allow for a non-vanishing CDM component by introducing $\omega_c$ as an additional free parameter. The remaining parameters are the same as in EDM 2p, and we continue to fix the transition sharpness.
Finally, in the most general case, EDM 4p, we also let the smoothing parameter $\sigma_\edm$ vary, making it a fully flexible model governed by four free parameters.
 Throughout our analysis, we follow the Planck conventions for neutrinos, and we include two massless
and one massive species with $m_\nu= 0.06$ eV~\cite{Planck:2018vyg}, and enforce $N_{\rm eff}=3.046$.

\section{Data analysis results}
\label{Sec:Results}

\begin{figure*}
	\begin{center}
    \includegraphics[width=\columnwidth]{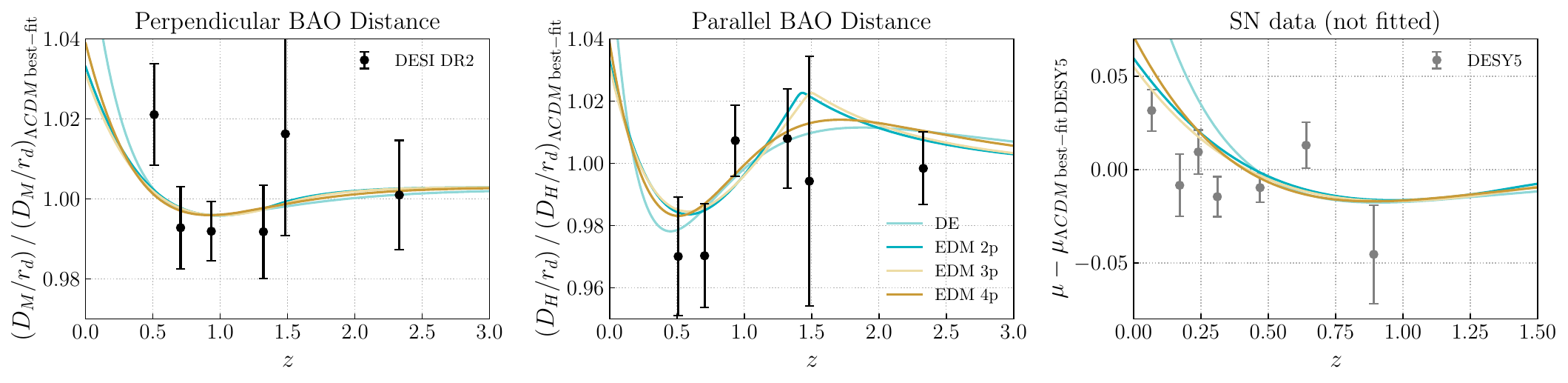}
    \includegraphics[width=\columnwidth]{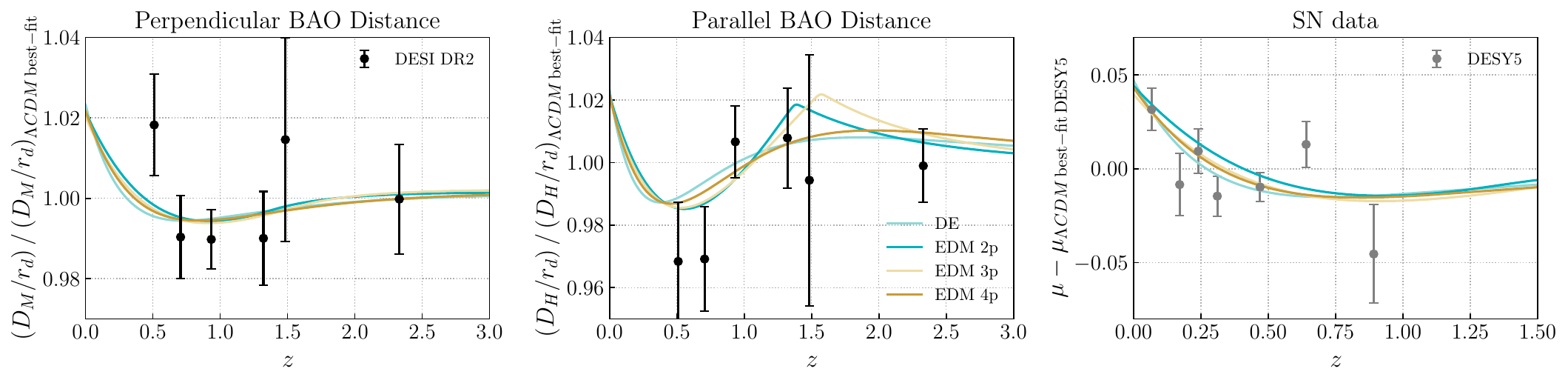}
    \includegraphics[width=\columnwidth]{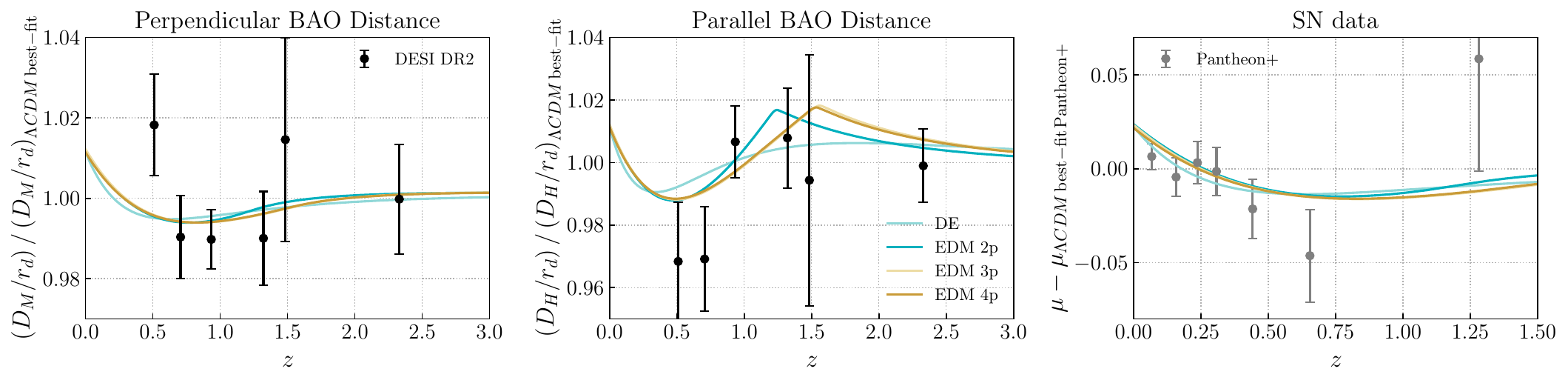}
    \includegraphics[width=\columnwidth]{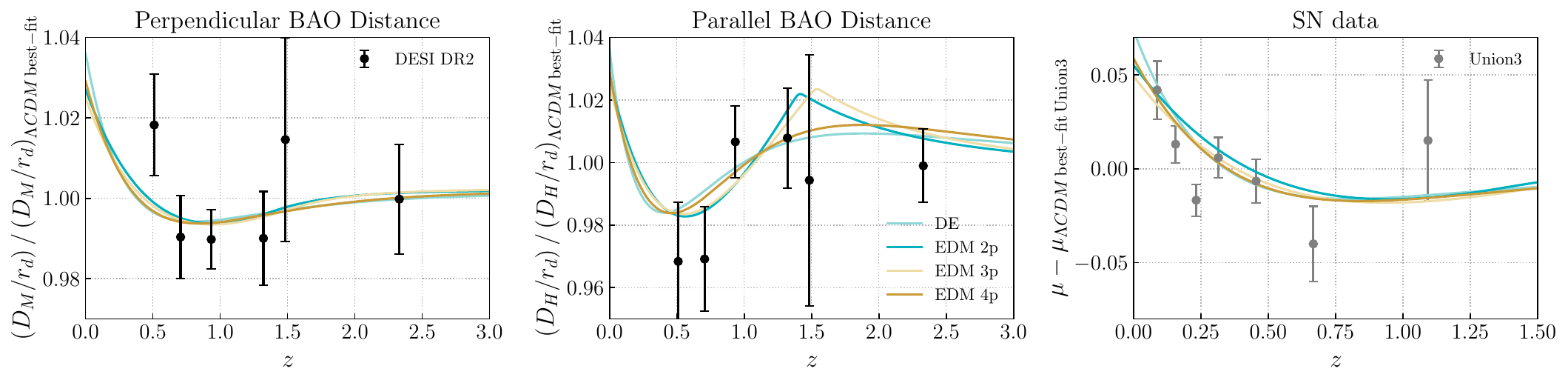}
	\end{center}\caption{\footnotesize\label{fig:best-fit} Hubble diagrams showing comparisons of DESI BAO and SNe data to the best-fit DE and EDM models compared to the $\Lambda$CDM models. 
From the left to the right we show the fit to DESI DR2 measurements of the perpendicular and parallel BAO distances, and supernovae distance modulus from DES-Y5. From the top to the bottom panels, we show the best-fits from the analysis with DESI DR2 + CMB, and then add DES-Y5, Pantheon+, or Union3 in this order. In the top-right panel, where SNe data are not employed in the analysis, the $\Lambda$CDM best-fit is taken to be that from the analysis including DES-Y5.
For visualization purposes SNe data are binned following the method described in~\cite{DESI:2025zgx}.
}
\end{figure*}

\begin{figure*}
	\begin{center}
		\includegraphics[width=.495\columnwidth]{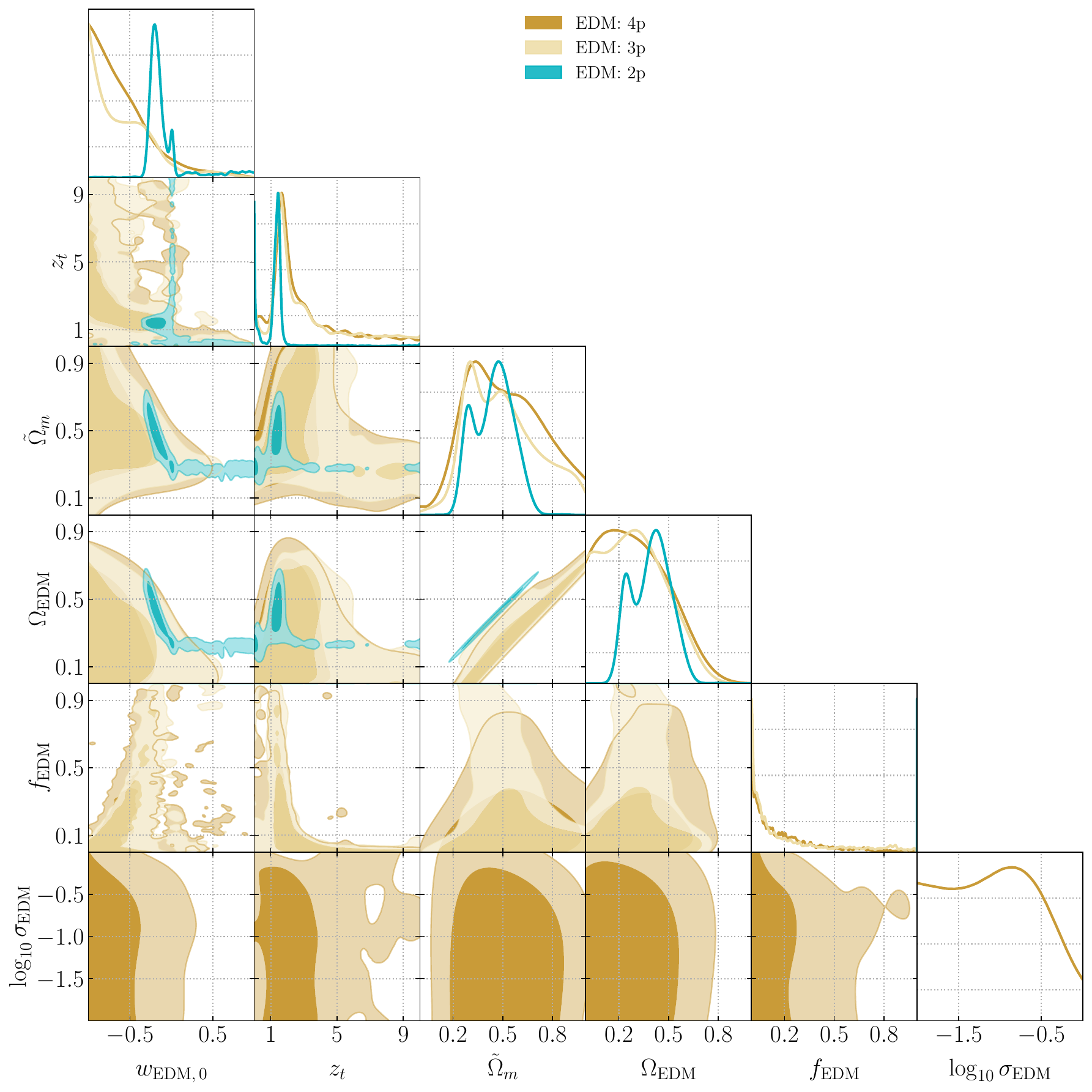}
		\includegraphics[width=.495\columnwidth]{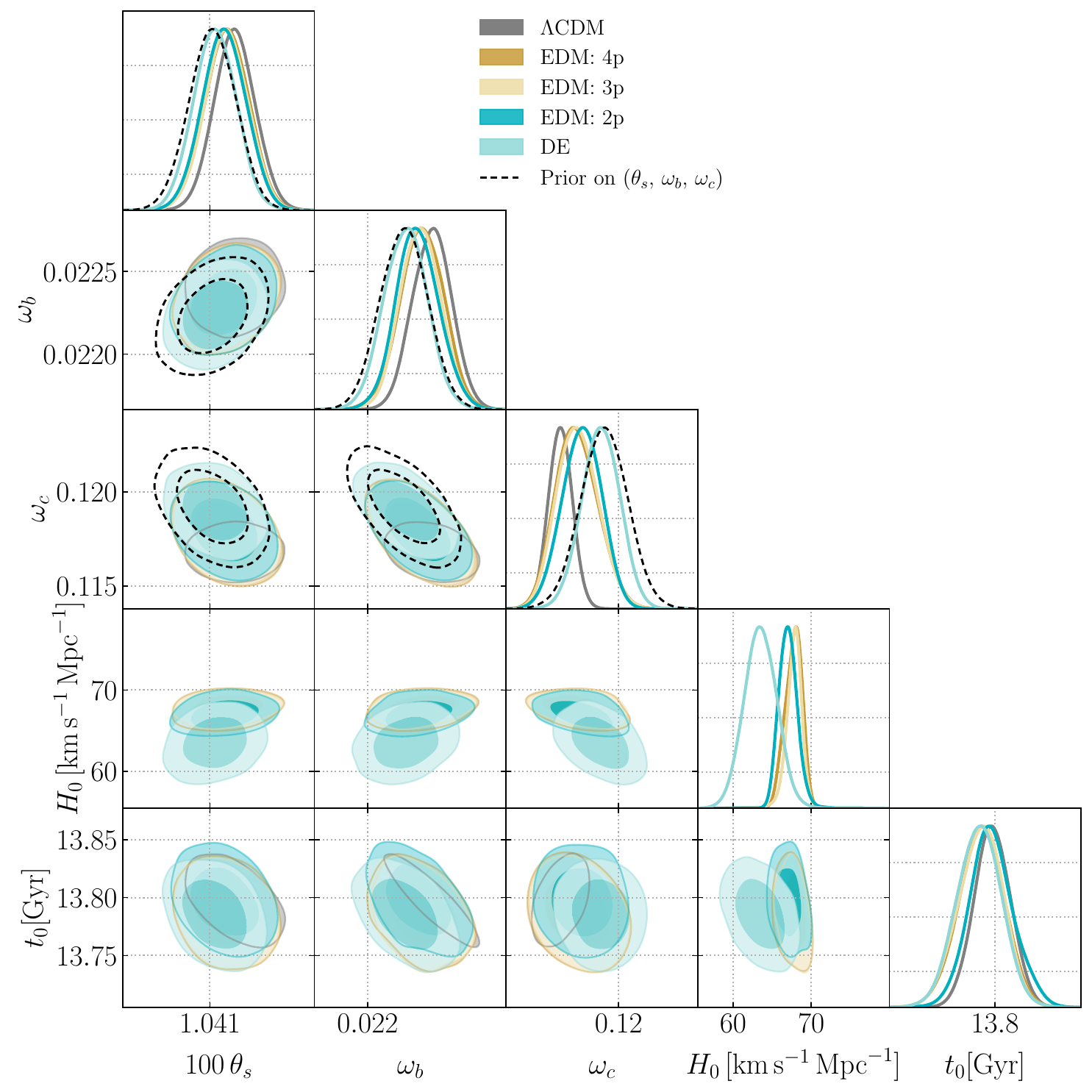}
        \includegraphics[width=\columnwidth]{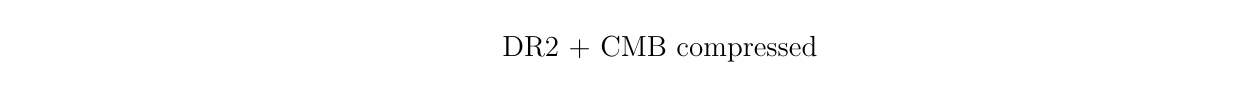}
	\end{center}\caption{\footnotesize\label{fig:constraints_DR2} 1D and 2D posterior distributions of the EDM parameters [left] and of the parameters common to $\Lambda$CDM, DE and EDM [right] from the analyses of the DR2 + CMB compressed dataset. In the second triangle plot, we also plot the prior on the parameters $(\theta_s,\,\omega_b,\,\omega_{c})$ for a comparison. }
\end{figure*}

\begin{table}[ht]
	\centering
		\small
	\begin{tabular}{ c c c c c c}
		\hline
		  & $\Lambda$CDM&DE & EDM 2p & EDM 3p  & EDM 4p  \\ 
		\hline
        $H_0$ &$68.33\pm 0.30$& $63.5\pm 2.1$ & $66.8^{+1.1}_{-1.3}$&$67.8^{+1.1}_{-0.83}$ &$67.7^{+1.2}_{-0.97}$\\
	&(68.33)&(63.91)&(66.26)&(66.40)&(65.91)\\
	$\omega_b*100$  & $2.241\pm 0.012$&$2.224\pm 0.014$ & $2.230^{+0.013}_{-0.014}$&$2.233\pm 0.014$ &$2.234\pm 0.014$\\
	&(2.241)&(2.226)&(2.228)&(2.227)&(2.227)\\
	$\omega_c*100$  &$11.679\pm 0.065$ &  $11.90\pm 0.10$&&$9.79^{+2.4}_{-0.53}$ &$9.86^{+2.2}_{-0.67}$\\
	&(11.68)&(11.89)&&(5.11)&(0.60)\\
        $w_0$ &-&$-0.41^{+0.21}_{-0.26}$  &-&-&- \\
        &&(-0.48)&&&\\
        $w_a$  &-&$-1.75^{+0.78}_{-0.60}$ &-&-&- \\
        &&(-1.52)&&&\\
        $w_{\edm,\,0}$ &-&-&$-0.16^{+0.69}_{-0.22}$(3$\sigma$)&$< 0.391$ (3$\sigma$)&$< 0.659$ (3$\sigma$)\\
        &&&(-0.21)&(-0.33)&(-0.36)\\
        $z_t$ &-&-&Unconstrained&Unconstrained& Unconstrained\\
        &&&(1.42)&(1.49)&(1.27)\\
        $\tilde{\Omega}_m$ &$0.2996\pm 0.0038$&$0.353^{+0.022}_{-0.026}$& $0.43^{+0.32}_{-0.29}$ (3$\sigma$) & $> 0.0884$ (3$\sigma$) & $0.51^{+0.44}_{-0.35}$ (3$\sigma$)  \\
        &&&(0.52)&(0.56)&(0.70)\\
        $\log_{10}\sigma_{\edm}$&-&-&-&-& Unconstrained \\ 
        &&&&&(-0.55)\\\hline
        $\theta_{s,\,{\rm red}}$ & $28\pm 20$& $6\pm 24$  &$15\pm 24$  &$18\pm 24$& $19\pm 25$ \\
        $t_0\,[\mathrm{Gyr}] $  &$13.798\pm 0.016$  & $13.786\pm 0.020$ &$13.794\pm 0.020$   &$13.788\pm 0.020$ &$13.788\pm 0.020$   \\
        $\Omega_{\edm}$ &-&-&$0.43^{+0.32}_{-0.29}$ (3$\sigma$) & $<0.73$ (3$\sigma$)  & $<0.80$ (3$\sigma$) \\
        $f_{\edm}$&-&-&-&$<0.98$ (3$\sigma$) &  $<0.94$ (3$\sigma$)\\
        \hline
		$\Delta \chi^2_{\rm DESI\, DR2}$  &-&4.2 &-3.5&-3.5 &-3.8\\\
		$\Delta \chi^2_{\rm CMB\, prior}$&-& -3.4&-3.2& -3.3&-3.3\\
		$\Delta \chi^2_{\rm total}$&-&-7.6 &-6.7&-6.8 &-7.1\\\hline
        ${\Delta{\rm AIC}}$&-&-3.6&-2.7&-0.8&+0.9\\\hline
		$\ln B$ &-&$-0.07\pm0.26$ &$+2.65\pm0.23$ & $+0.86\pm0.22$&$+0.84\pm0.23$\\
		 &&Bare & Strongly &Barely &Barely \\
		 &&mention &disfavored &disfavored&disfavored \\\hline
	\end{tabular}
	
	\caption{Constraints on main and derived parameters considering the data set DR2 + CMB
		for $\Lambda$CDM, DE, and the EDM models with 2, 3 ad 4 parameters. Unless stated otherwise, we report mean values and the 68\% CL (1$\sigma$). We also report $\Delta\chi^2$, 
        ${\Delta{\rm AIC}}$ and Bayes factors of the extended models with respect to the $\Lambda$CDM model. We remind the reader that $\tilde{\Omega}_m\equiv\Omega_b+\Omega_\cdm+\Omega_\edm$ is a primary parameter in the EDM analysis, whereas it is a derived parameter in the other two models, and that $f_{\rm EDM}$ is fixed to 1 in the EDM 2p model. For reasons of space we define $\theta_{s,\,{\rm red}}\equiv(100*\theta_{s}-1.041)*10^5$. Best-fit primary parameters are reported inside the parenthesis.
        }
	\label{tab:constraints_DR2}
\end{table}

\begin{figure*}
	\begin{center}
		\includegraphics[width=.495\columnwidth]{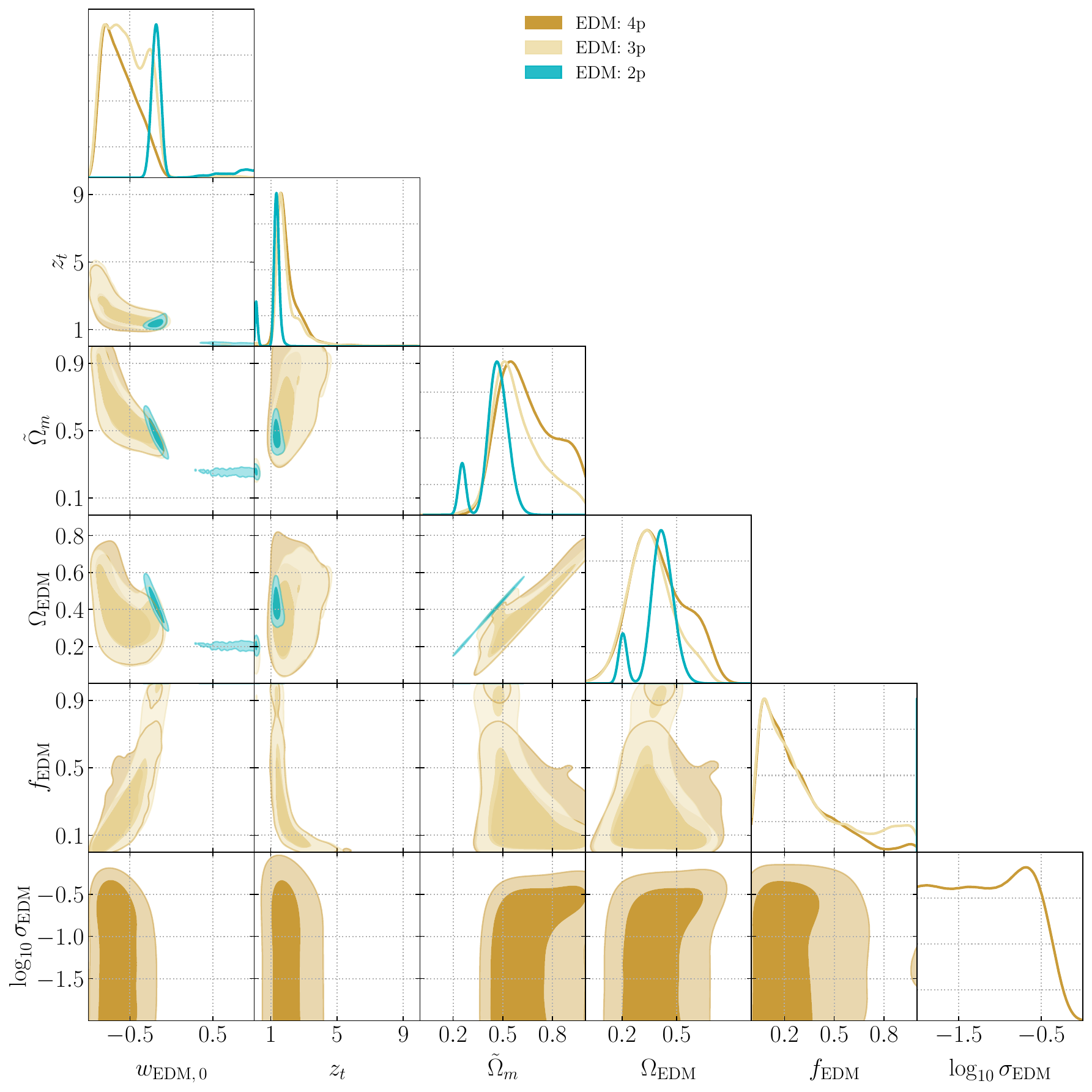}
		\includegraphics[width=.495\columnwidth]{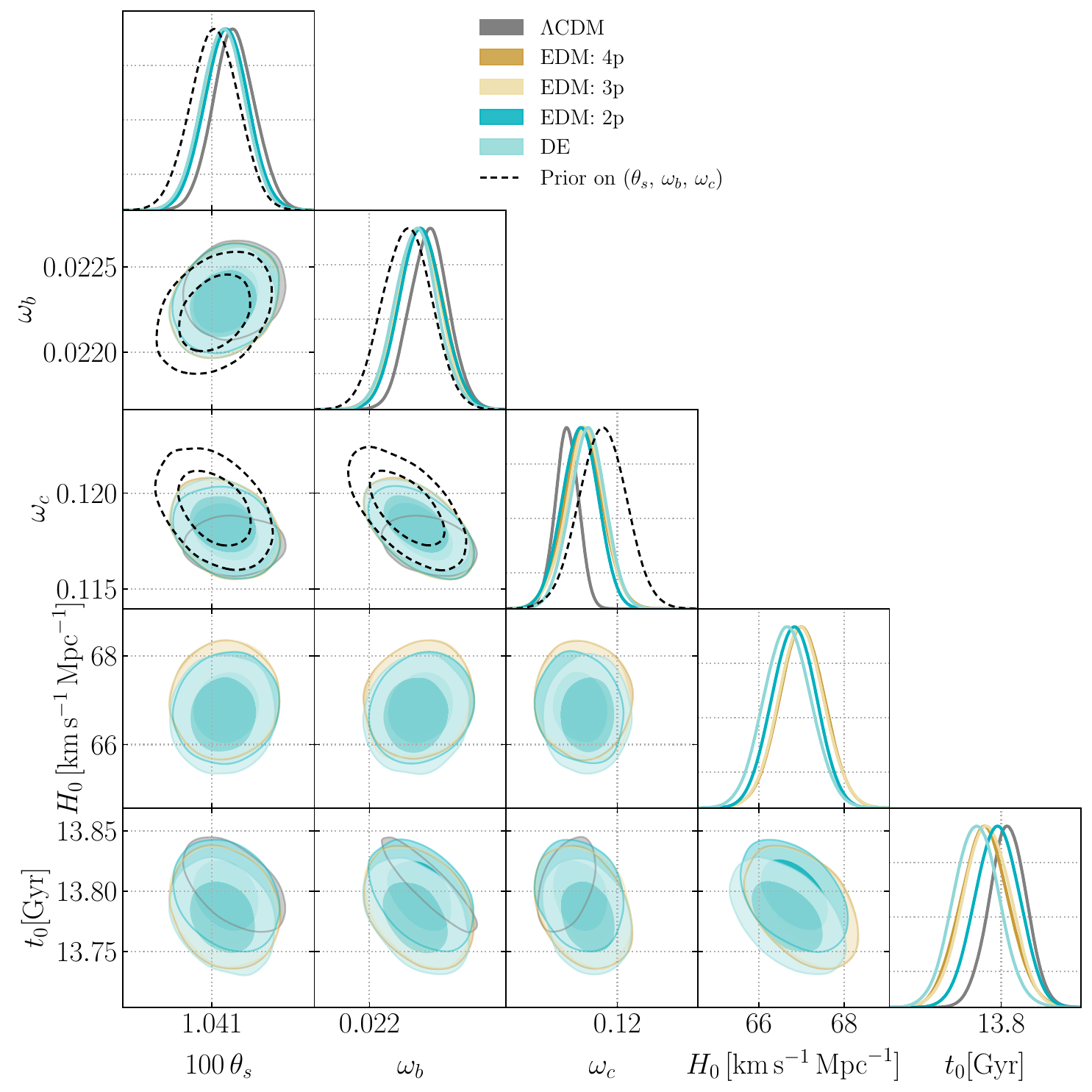}
        \includegraphics[width=\columnwidth]{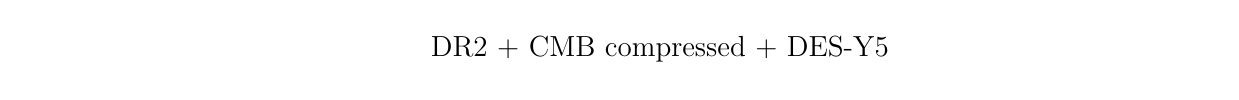}
	\end{center}\caption{\footnotesize\label{fig:constraints_DR2_D5} 1D and 2D posterior distributions of the EDM parameters [left] and of the parameters common to $\Lambda$CDM, DE and EDM [right] from the analyses of the DR2 + CMB compressed + DES-Y5 dataset. In the second triangle plot, we also plot the prior on the parameters $(\theta_s,\,\omega_b,\,\omega_{c})$ for a comparison. }
\end{figure*}

\begin{table}[ht]
	\centering
	
	\begin{tabular}{ c c c c c c}
		\hline
		& $\Lambda$CDM&DE & EDM 2p & EDM 3p  & EDM 4p  \\ 
		\hline
		$H_0$ &$68.10\pm 0.29$ & $66.67\pm 0.55$&$66.83\pm 0.51$& $66.93\pm 0.52$&$67.01\pm 0.54$\\
		&(68.11)&(66.65)&(66.74)&(66.88)&(66.77)\\
		$\omega_b*100$  & $2.236\pm 0.012$ &$2.228\pm 0.013$ &$2.231\pm 0.013$&$2.230\pm 0.013$ &$2.231\pm 0.014$\\
		&(2.236)&(2.229)&(2.231)&(2.226)&(2.228)\\
		$\omega_c*100$  &$11.727\pm 0.064$& $11.838\pm 0.097$&&$8.22^{+3.4}_{-0.74}$ &$8.87^{+2.6}_{-0.77}$\\
		&(11.70)&(11.83)&&(8.15)&(6.60)\\
		$w_0$ &-&$-0.763\pm 0.057$&-&-&- \\
		&&(-0.76)&&&\\
		$w_a$  &-&$-0.79^{+0.25}_{-0.22}$ &-&-&- \\
		&&(-0.80)&&&\\
		$w_{\edm,\,0}$ &-&-&$ > -0.307$ (3$\sigma$)&$-0.55\pm 0.22$&$-0.61^{+0.13}_{-0.27}$\\
		&&&(-0.18)&(-0.49)&(-0.72)\\
		$z_t$ &-&-&$<1.92$ (3$\sigma$)&$2.06^{+0.11}_{-0.83}$&$2.08^{+0.19}_{-0.85}$ \\
		&&&(1.38)&(1.57)&(1.40)\\
		$\tilde{\Omega}_m$ & $0.3025\pm 0.0037$&$0.3180\pm 0.0056$& $0.461^{+0.075}_{-0.051}$ & $0.609^{+0.091}_{-0.19}$&  $0.65^{+0.13}_{-0.22}$  \\
		&&&(0.47)&(0.57)&(0.98)\\
		$\log_{10}\sigma_{\edm}$&-&-&-&-& $< -0.280$ (3$\sigma$)  \\ 
		&&&&&(-0.37)\\\hline
		$\theta_{s,\,{\rm red}}$ &$24\pm 20$ & $12\pm 24$ & $16\pm 24$ &$14\pm 24$ & $15\pm 24$ \\
		$t_0\,[\mathrm{Gyr}] $  &$13.806\pm 0.016$  & $13.779\pm 0.020$ & $13.795\pm 0.019$  &$13.787^{+0.020}_{-0.018}$ & $13.786\pm 0.021$ \\
		$\Omega_{\edm}$ &-&-& $0.411^{+0.075}_{-0.051}$ & $0.38^{+0.11}_{-0.15}$& $0.40^{+0.13}_{-0.19}$  \\
		$f_{\edm}$&-&-&-& $0.304^{+0.069}_{-0.29}$&  $0.249^{+0.071}_{-0.23}$\\
\hline
$\Delta \chi^2_{\rm DESI\, DR2}$  &-&-4.4 &-5.3& -5.1&-4.9\\
$\Delta \chi^2_{\rm CMB\, prior}$  &-&-1.8 &-1.6& -2.1&-2.0\\
$\Delta \chi^2_{\rm DES-Y5}$  &-&-11.8 &-10.2&-10.5 &-11.1\\
$\Delta \chi^2_{\rm total}$ &-& -18.0&-17.1& -17.7&-18.0 \\\hline	
        ${\Delta{\rm AIC}}$&-&-14.0&-13.1&-11.7&-10.0\\\hline
$\ln B$ &-&$-4.29\pm0.26$ &$-1.68\pm0.23$&$-3.34\pm0.23$ &$-3.19\pm0.23$ \\
&&Very strong &Substantial & Strong & Strong\\\hline\end{tabular}
	
	\caption{Constraints on main and derived parameters considering the data set DR2 + CMB + DES-Y5
		for $\Lambda$CDM, DE, and the EDM models with 2, 3 ad 4 parameters. Unless stated otherwise, we report mean values and the 68\% CL (1$\sigma$). We also report $\Delta\chi^2$, 
        ${\Delta{\rm AIC}}$ and Bayes factors of the extended models with respect to the $\Lambda$CDM model. We remind the reader that $\tilde{\Omega}_m\equiv\Omega_b+\Omega_\cdm+\Omega_\edm$ is a primary parameter in the EDM analysis, whereas it is a derived parameter in the other two models, and that $f_{\rm EDM}$ is fixed to 1 in the EDM 2p model. Best-fit primary parameters are reported inside the parenthesis. }
	\label{tab:constraints_D5}
\end{table}

\begin{figure*}
	\begin{center}
		\includegraphics[width=.495\columnwidth]{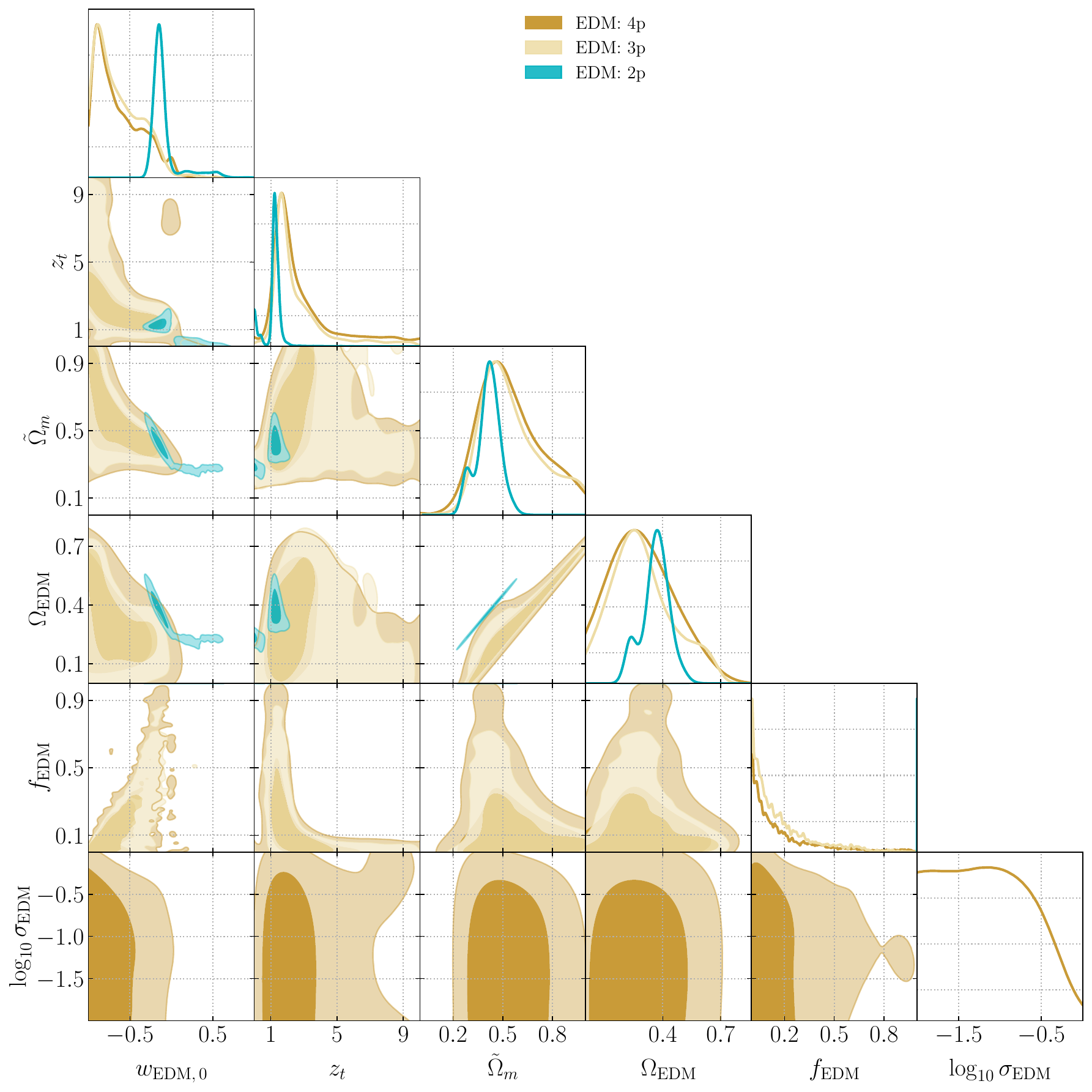}
		\includegraphics[width=.495\columnwidth]{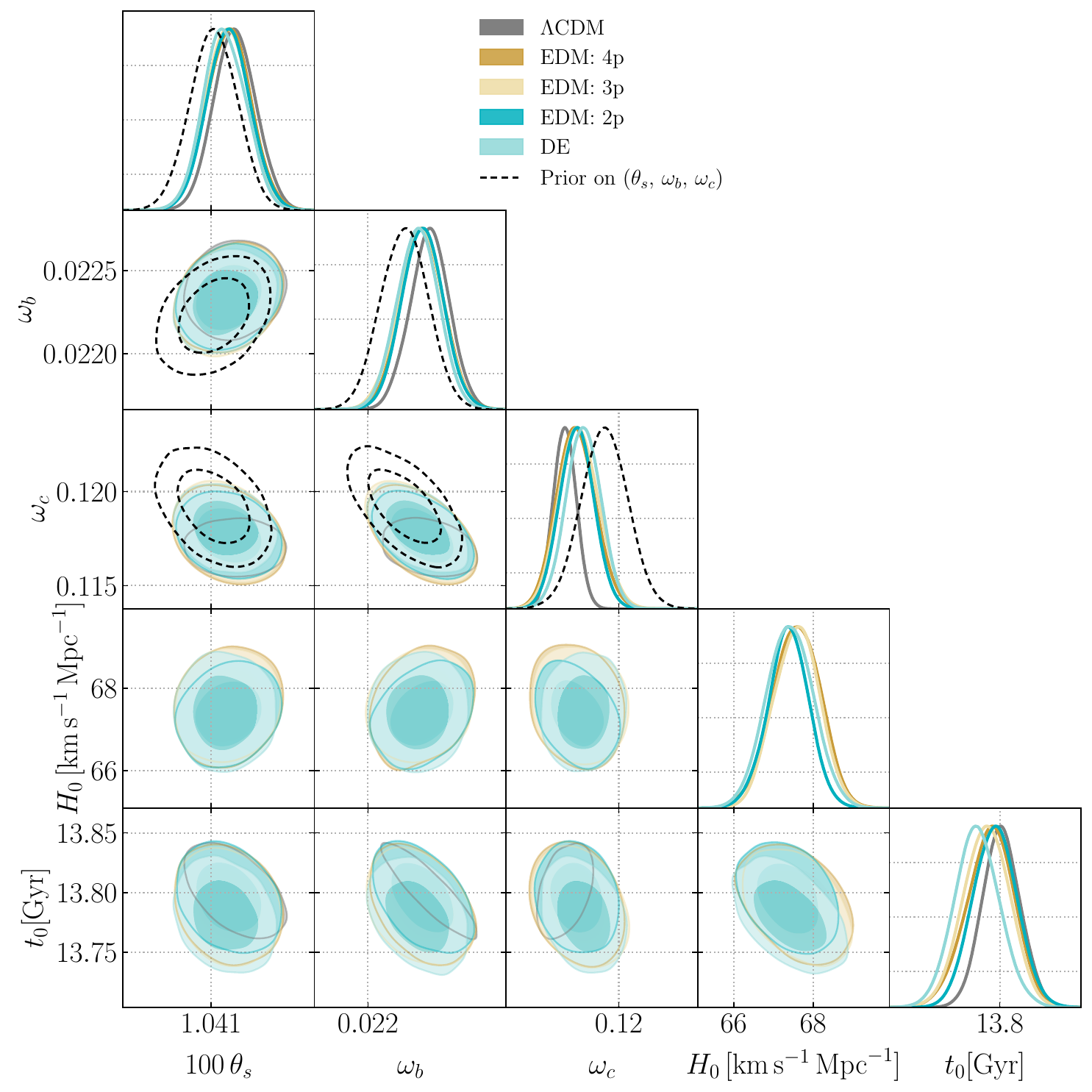}
        \includegraphics[width=\columnwidth]{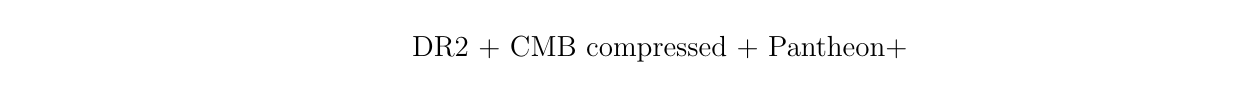}
	\end{center}\caption{\footnotesize\label{fig:constraints_DR2_P} 1D and 2D posterior distributions of the EDM parameters [left] and of the parameters common to $\Lambda$CDM, DE and EDM [right] from the analyses of the DR2 + CMB compressed + Pantheon+ dataset. In the second triangle plot, we also plot the prior on the parameters $(\theta_s,\,\omega_b,\,\omega_{c})$ for a comparison. }
\end{figure*}

\begin{table}[ht]
	\centering
	
	\begin{tabular}{ c c c c c c}
		\hline
		& $\Lambda$CDM&DE & EDM 2p & EDM 3p  & EDM 4p  \\ 
		\hline
		$H_0$ &$68.22\pm 0.30$&$67.42\pm 0.59$ &$67.38\pm 0.56$&$67.61\pm 0.53$ &$67.58\pm 0.60$\\
		&(68.22)&(67.41)&(67.36)&(67.38)&(67.43)\\
		$\omega_b*100$  &$2.239\pm 0.012$&$2.231\pm 0.013$  &$2.235\pm 0.013$& $2.233\pm 0.013$&$2.234\pm 0.013$ \\
		&(2.239)&(2.232)&(2.231)&(2.226)&(2.230)\\
		$\omega_c*100$  &$11.702\pm 0.064$&$11.805\pm 0.096$&&$9.78^{+2.0}_{-0.30}$ & $9.76^{+2.1}_{-0.30}$\\
		&(11.70)&(11.80)&&(9.09)&(9.04)\\
		$w_0$ &-&$-0.850\pm 0.055$&-&-&- \\
		&&(-0.85)&&&\\
		$w_a$  &-&$-0.53^{+0.22}_{-0.20}$&-&-&- \\
		&&(-0.51)&&&\\
		$w_{\edm,\,0}$ &-&-&$-0.11^{+0.85}_{-0.19}$ (3$\sigma$)&$< 0.0562$ (3$\sigma$)&$< 0.0807$ (3$\sigma$)\\
		&&&(-0.17)&(-0.53)&(-0.52)\\
		$z_t$ &-&-&$< 6.43$ (3$\sigma$)&$< 8.94$ (3$\sigma$) & Unconstrained\\
		&&&(1.23)&(1.55)&(1.53)\\
		$\tilde{\Omega}_m$ &$0.3009\pm 0.0037$&$0.3103\pm 0.0057$&$0.41^{+0.16}_{-0.19}$ (3$\sigma$) & $ > 0.259$ (3$\sigma$) & $> 0.230$ (3$\sigma$)  \\
		&&&(0.44)&(0.52)&(0.51)\\
		$\log_{10}\sigma_{\edm}$&-&-&-&-& Unconstrained \\ 
		&&&&&(-1.97)\\\hline
		$\theta_{s,\,{\rm red}}$ & $26\pm 20$ &$15\pm 24$& $20\pm 20$ &$18\pm 24$&  $20\pm 25$ \\
		$t_0\,[\mathrm{Gyr}] $  & $13.801\pm 0.016$ & $13.780\pm 0.020$& $13.796\pm 0.019$  &$13.788\pm 0.020$ & $13.792^{+0.022}_{-0.019}$ \\
		$\Omega_{\edm}$ &-&-&$0.36^{+0.16}_{-0.19}$ (3$\sigma$) & $<0.71$ (3$\sigma$) & $<0.69$ (3$\sigma$) \\
		$f_{\edm}$&-&-&-& $<0.86$ (3$\sigma$) & $<0.95$ (3$\sigma$) \\
		\hline
		$\Delta \chi^2_{\rm DESI\, DR2}$  &-&-2.6 &-3.7&-3.8 &-3.8\\
		$\Delta \chi^2_{\rm CMB\, prior}$  &-&-2.0 &-2.0&-2.5 &-2.4\\
		$\Delta \chi^2_{\rm Phanteon+}$  &-&-2.9 &-3.4&-3.2 &-3.3\\
		$\Delta \chi^2_{\rm total}$ &-&-7.5 &-9.1&-9.5 &-9.5 \\\hline
        ${\Delta{\rm AIC}}$&-&-3.5&-5.1&-3.5&-1.5\\\hline
		$\ln B$&-&$+1.38\pm0.27$ &$+1.85\pm0.23$& $+0.02\pm0.23$&$+0.21\pm0.23$\\
		&&Substantially &Substantially & Barely& Barely\\
		&&disfavored &disfavored & disfavored& disfavored\\\hline
	\end{tabular}
	
	\caption{Constraints on main and derived parameters considering the data set DR2 + CMB + Pantheon+
		for $\Lambda$CDM, DE, and the EDM models with 2, 3 ad 4 parameters. Unless stated otherwise, we report mean values and the 68\% CL (1$\sigma$). We also report $\Delta\chi^2$, 
        ${\Delta{\rm AIC}}$ and Bayes factors of the extended models with respect to the $\Lambda$CDM model. We remind the reader that $\tilde{\Omega}_m\equiv\Omega_b+\Omega_\cdm+\Omega_\edm$ is a primary parameter in the EDM analysis, whereas it is a derived parameter in the other two models, and that $f_{\rm EDM}$ is fixed to 1 in the EDM 2p model.  Best-fit primary parameters are reported inside the parenthesis.}
	\label{tab:constraints_P+}
\end{table}
\begin{figure*}
	\begin{center}
		\includegraphics[width=.495\columnwidth]{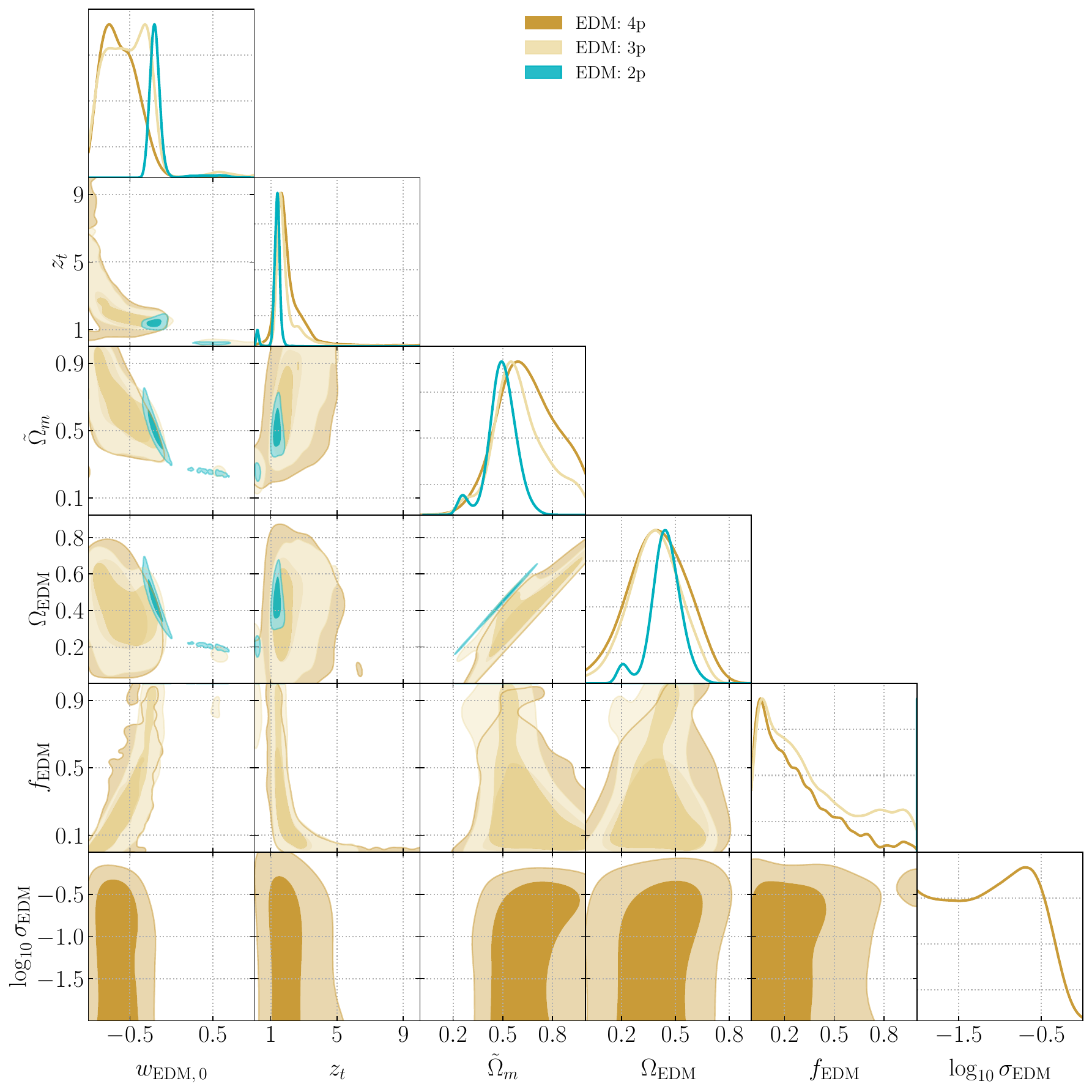}	\includegraphics[width=.495\columnwidth]{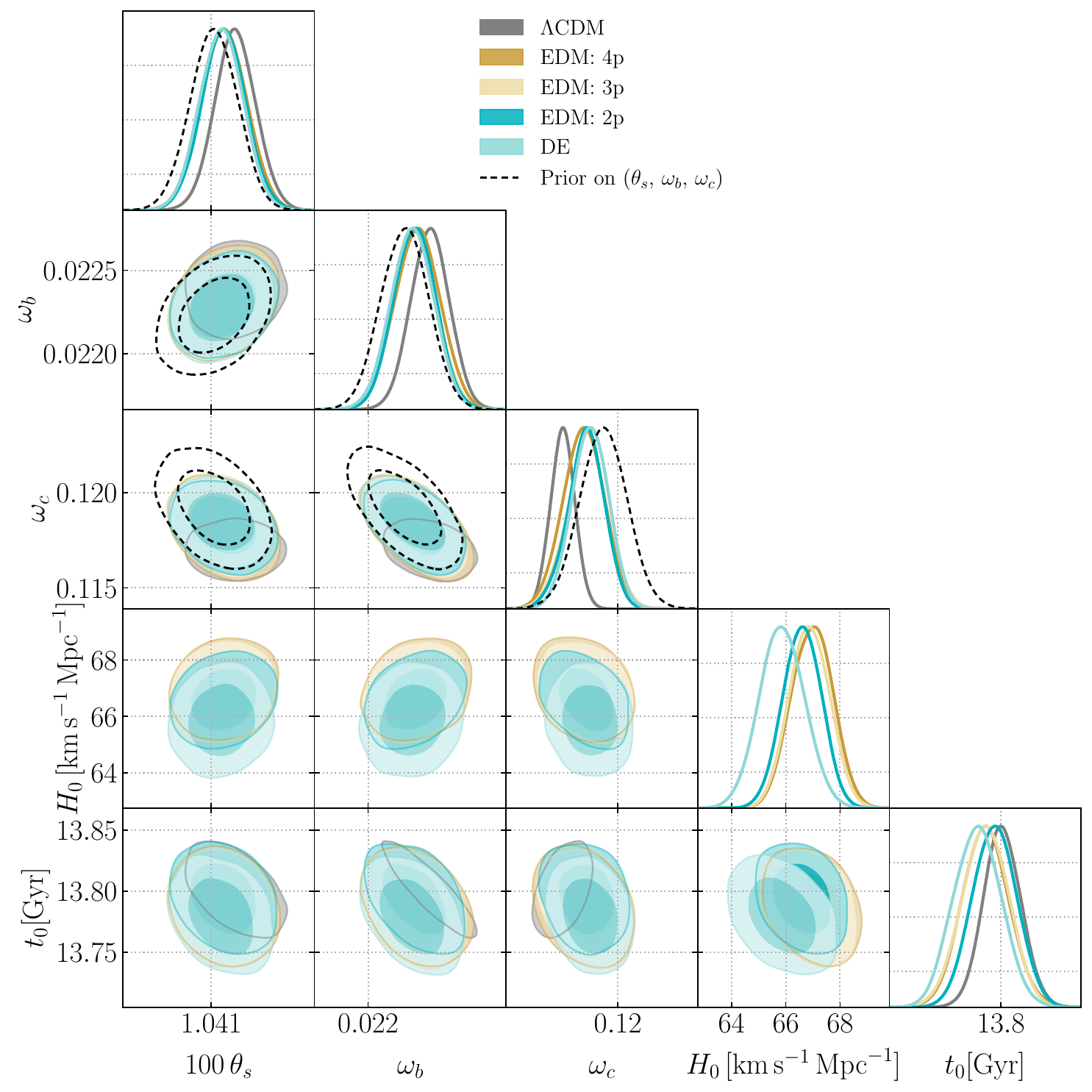}

        \includegraphics[width=\columnwidth]{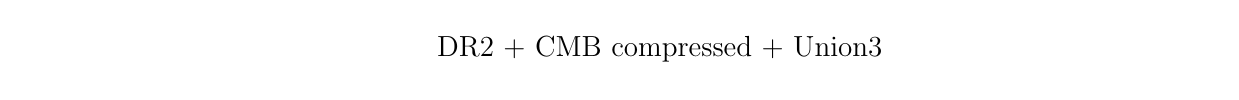}	
	\end{center}\caption{\footnotesize\label{fig:constraints_DR2_U} 1D and 2D posterior distributions of the EDM parameters [left] and of the parameters common to $\Lambda$CDM, DE and EDM [right] from the analyses of the DR2 + CMB compressed + Union3 dataset. In the second triangle plot, we also plot the prior on the parameters $(\theta_s,\,\omega_b,\,\omega_{c})$ for a comparison. }
\end{figure*}

\begin{table}[ht]
	\centering
	
	\begin{tabular}{ c c c c c c}
		\hline
		& $\Lambda$CDM&DE & EDM 2p & EDM 3p  & EDM 4p  \\ 
		\hline
		$H_0$ &$68.23\pm 0.30$& $65.87\pm 0.83$&$66.53\pm 0.68$& $66.98\pm 0.74$&$66.97\pm 0.76$ \\
		&(68.23)&(65.83)&(66.40)&(66.58)&(66.29)\\
		$\omega_b*100$  &$2.239\pm 0.012$& $2.228\pm 0.013$&$2.230\pm 0.013$& $2.229\pm 0.015$ & $2.231\pm 0.014$\\
		&(2.239)&(2.228)&(2.228)&(2.226)&(2.227)\\
		$\omega_c*100$  &$11.699\pm 0.066$&$11.849\pm 0.099$ &&$8.71^{+3.0}_{-0.96}$ &$8.80^{+2.9}_{-0.97}$\\
		&(11.70)&(11.85)&&(6.85)&(3.44)\\
		$w_0$ &-&$0.3261\pm 0.0087$ &-&-&- \\
		&&(-0.68)&&&\\
		$w_a$  &-&$-1.01\pm 0.31$ &-&-&- \\
		&&(-1.01)&&&\\
		$w_{\edm,\,0}$ &-&-&$> -0.325$ (3$\sigma$)&$-0.56^{+0.37}_{-0.42}$ (3$\sigma$)&$< -0.0944$ (3$\sigma$)\\
		&&&(-0.21)&(-0.42)&(-0.69)\\
		$z_t$ &-&-&$<1.98$ (3$\sigma$)&$<8.17$ (3$\sigma$)& Unconstrained \\
		&&&(1.41)&(1.53)&(1.23)\\
		$\tilde{\Omega}_m$ &$0.3008\pm 0.0039$&$0.3261\pm 0.0087$&$0.49^{+0.22}_{-0.28}$ (3$\sigma$) & $> 0.249$ (3$\sigma$) & $> 0.270$ (3$\sigma$)  \\
		&&&(0.51)&(0.57)&(0.94)\\
		$\log_{10}\sigma_{\edm}$&-&-&-&-& $< -0.211$ (3$\sigma$)  \\ 
		&&&&&(-0.39)\\\hline
		$\theta_{s,\,{\rm red}}$ &$27\pm 20$ & $11\pm 24$   & $14\pm 24$ &$14\pm 25$& $15\pm 25$ \\
		$t_0\,[\mathrm{Gyr}] $  & $13.801\pm 0.016$ & $13.780\pm 0.020$  & $13.794\pm 0.019$  & $13.787\pm 0.020$ & $13.787\pm 0.020$ \\
		$\Omega_{\edm}$ &-&-&$0.44^{+0.22}_{-0.28}$ (3$\sigma$) & $<0.70$ (3$\sigma$) & $<0.8$ (3$\sigma$) \\
		$f_{\edm}$&-&-&-& $<0.92$ (3$\sigma$) & $<0.95$ (3$\sigma$) \\
		\hline
		$\Delta \chi^2_{\rm DESI\, DR2}$  &-&-4.1 &-4.2& -4.1& -4.1\\
		$\Delta \chi^2_{\rm CMB\, prior}$  &-&-2.6 &-2.5& -2.8&-2.8\\
		$\Delta \chi^2_{\rm Union3}$ &-&-6.9 &-6.1&-6.2 &-6.7\\
		$\Delta \chi^2_{\rm total}$ &-&-13.6 &-12.8& -13.1&-13.6\\\hline
        ${\Delta{\rm AIC}}$&-&-9.6&-8.8&-7.1&-5.6\\\hline	
		$\ln B$ &-&$-2.10\pm0.26$ &$-0.05\pm0.23$& $-1.24\pm0.23$&$-1.22\pm0.23$\\
		&&Substantial&Bare&Substantial&Substantial\\
		&&&mention&  &\\\hline\end{tabular}
	
	\caption{Constraints on main and derived parameters considering the data set DR2 + CMB + Union3
		for $\Lambda$CDM, DE, and the EDM models with 2, 3 ad 4 parameters. We report mean values and the 68\% CL, except for the case of upper or lower limits, for which we report the 95\% CL. We also report $\Delta\chi^2$, 
        ${\Delta{\rm AIC}}$ and Bayes factors of the extended models with respect to the $\Lambda$CDM model. We remind the reader that $\tilde{\Omega}_m\equiv\Omega_b+\Omega_\cdm+\Omega_\edm$ is a primary parameter in the EDM analysis, whereas it is a derived parameter in the other two models, and that $f_{\rm EDM}$ is fixed to 1 in the EDM 2p model. Best-fit primary parameters are reported inside the parenthesis. }
	\label{tab:constraints_U3}
\end{table}

\begin{figure*}
	\begin{center}
    \includegraphics[width=.45\columnwidth]{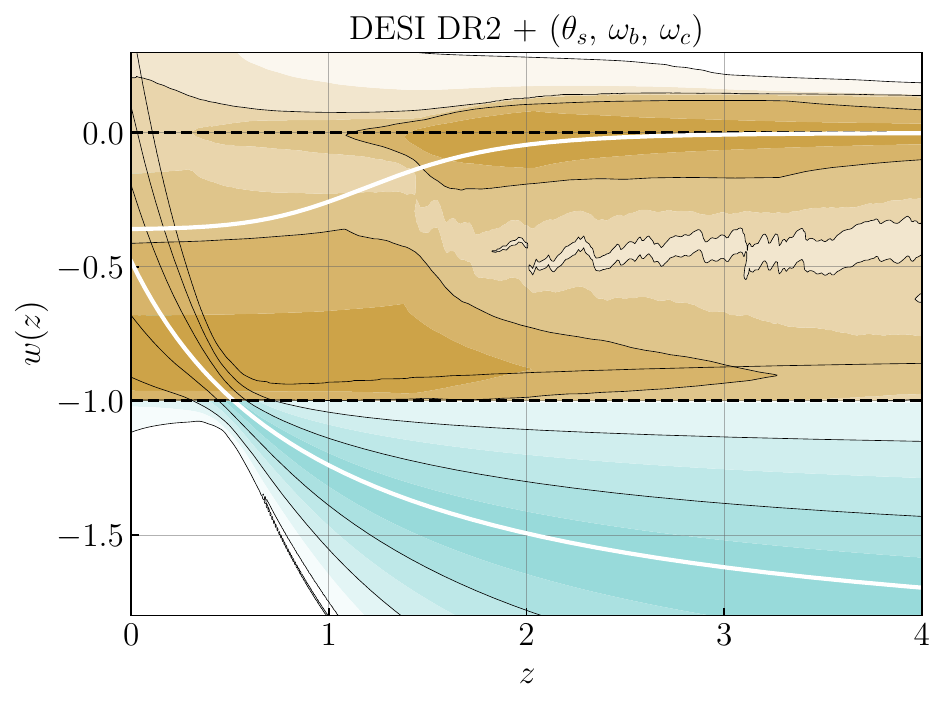}
    \includegraphics[width=.45\columnwidth]{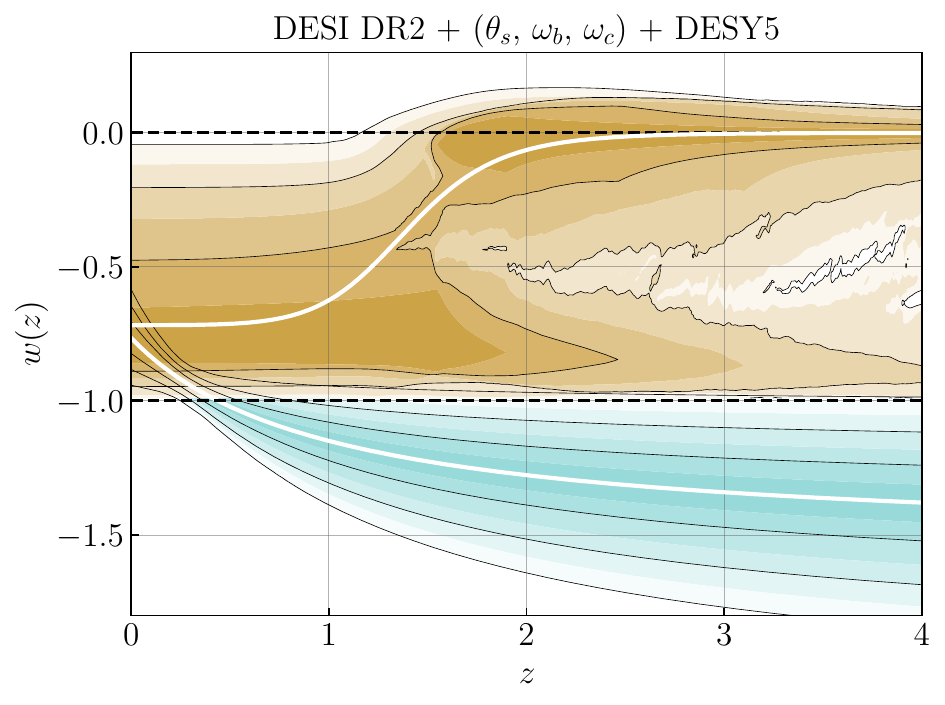}
    \includegraphics[width=.45\columnwidth]{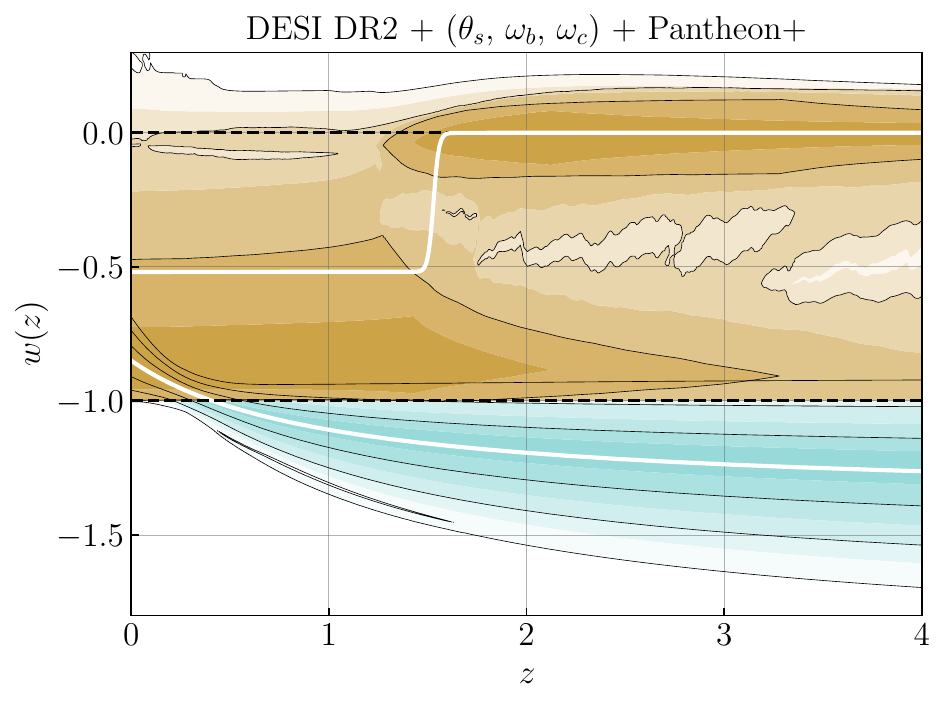}
    \includegraphics[width=.45\columnwidth]{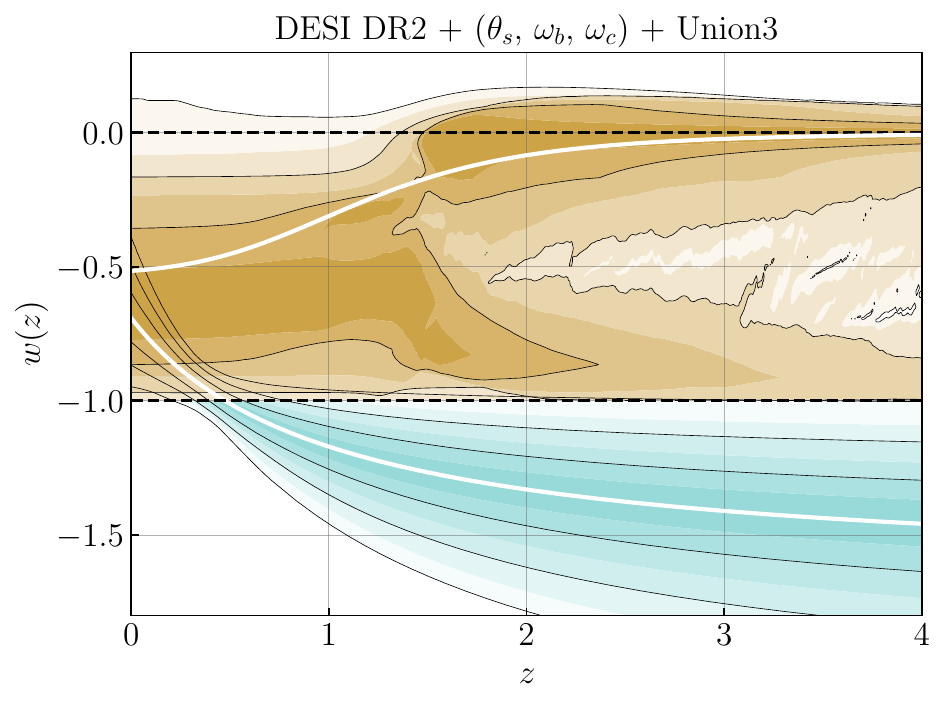}
    \includegraphics[width=\columnwidth]{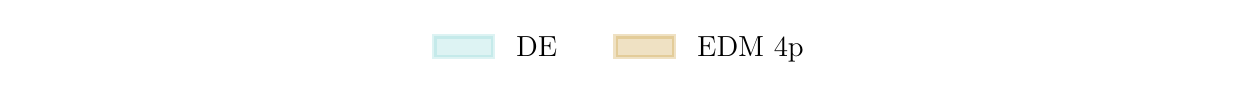}

	\end{center}\caption{\footnotesize\label{fig:EOS}Predictive posterior distributions for the EOS in the DE and EDM models, obtained from the analysis of the three datasets. For the EDM model, we show results from the analysis where the smoothing parameter $\sigma_{\rm EDM}$ is varied. Shading from darker to lighter indicates confidence levels from 68\% (1$\sigma$) to 99\% (3$\sigma$). The white lines indicate the best-fit EOS for each model. {Comparing this figure with Fig.~\ref{fig:3-para-models} and Fig.~\ref{fig:best-fit}, we see that, in producing similar background evolution, the behaviors of the EOS in the DE and EDM models are dramatically different, and, in particular, there is no phantom crossing for EDM.}
}
\end{figure*}

We are now in a position to discuss the results of our analysis. We present the results for each dataset in turn, in order to highlight the impact of adding the three different SNe samples to our baseline. For each dataset, we provide the parameter constraints, discuss model selection, and present the best-fit solutions for each model.

\paragraph{DR2 + CMB.} We present the constraints from the analysis of this dataset in Fig.~\ref{fig:constraints_DR2} and summarize them in Table~\ref{tab:constraints_DR2}. The best-fit predictions for the extended models are compared to the $\Lambda$CDM best-fit in the first row of Fig.~\ref{fig:best-fit}. A visual inspection of the best-fit models in Fig.~\ref{fig:best-fit} confirms our argument from Sec.~\ref{Sec:EDM models}: EDM models are capable of reproducing late-time dynamics that resemble those of evolving DE.

Locally in the parameter space, i.e.~regarding the best-fit models, we observe that the EDM models provide better fits to the dataset than $\Lambda$CDM, although they perform slightly worse than the DE model. In particular, while the 3p and 4p EDM models (but not the 2p one) yield a marginally better fit to the DR2 data compared to DE, this improvement comes at the cost of a degraded fit to the CMB compressed likelihood.

Globally in the entire explored parameter space, although the posterior distributions of the EDM models clearly peak (as seen in the left panel of Fig.~\ref{fig:constraints_DR2}), the significance of these peaks is only at the $\sim1\sigma$ level, with no statistically significant detection of any EDM parameter. For example, for all three EDM models the prominent 68\%CL mode in the posterior of $z_t$ becomes a bound at 95\%CL, and the parameter is unconstrained at 99\%CL due to the long tail extending to large values of $z_t$. Due to this and the larger number of free parameters, EDM is slightly disfavored from a model selection perspective according to Jeffreys’ scale~\cite{Kass:1995loi}, with the 2p model being the most disfavored. While this may seem counterintuitive, the reason is straightforward: the 2p model lacks the flexibility of the 3p and 4p versions, and as seen in Fig.~\ref{fig:constraints_DR2}, a much narrower parameter volume is consistent with data, in addition to providing a poorer fit. This wasted prior volume penalizes the Bayesian evidence.

We also note that both DE and EDM are consistent with the standard model, as shown in Fig.~\ref{fig:EOS}. Indeed, for the DE model, a constant EOS with $w_\Lambda=-1$ is consistent with the predictive posterior distribution of $w_{\rm DE}(z)$ at 3$\sigma$ (but not at 2$\sigma$). For the EDM models, while we see a $\sim1.5\sigma$ hint of $w_{\rm EDM}(z)\neq0$ at $z\lesssim1$, this is consistent with $w_\edm=w_{\rm CDM}=0$ at less than 2$\sigma$. 

Despite this, it is still informative to consider the physical properties of the best-fit solutions. All models tend to favor negative values of the EDM's EOS, implying that the EDM component redshifts more slowly than standard matter. This, in turn, leads to a higher total value of $\tilde{\Omega}_m$, as anticipated, pushing the matter–cosmological constant equality into the future. We will see that these features persist in the best-fit solutions obtained from the extended datasets as well.

Regarding the standard cosmological parameters, we find that in the EDM models they remain broadly consistent with the estimates obtained under $\Lambda$CDM and DE. Interestingly, the mean value of the present-day Hubble parameter, $H_0$, in EDM is slightly lower than in $\Lambda$CDM. However, the associated uncertainties are nearly four times larger, which results in a mild alleviation of the $H_0$ tension. While the effect is modest, it is nonetheless noteworthy that, in contrast, the DE model tends to exacerbate the tension: despite also yielding larger error bars, it shifts the mean $H_0$ value downward, thereby worsening the discrepancy with local measurements. We also estimated the age of the Universe $t_0$ in all the models, and we can see that the differences are not large enough to discriminate the models using current astrophysical data~\cite{Vagnozzi:2021tjv}.

\paragraph{DR2 + CMB + DES-Y5.} We present the constraints from the analysis of this dataset in Fig.~\ref{fig:constraints_DR2_D5} and summarize them in Table~\ref{tab:constraints_D5}. The best-fit predictions for the extended models are compared to the $\Lambda$CDM best fit in the second row of Fig.~\ref{fig:best-fit}. 

This dataset is well known for providing the strongest evidence in favor of an evolving DE fluid, yielding a $3.25\sigma$ detection of the parameter $w_a$ if we approximate its posterior as a Gaussian, indicative of a time-dependent DE EOS. Furthermore, the present-day EOS deviates from $-1$ at $\gtrsim4\sigma$ level. The dynamical nature of DE is clearly illustrated in the upper-right panel of Fig.~\ref{fig:EOS}, where a so-called phantom crossing is visible --- i.e., the EOS dips below $-1$ at intermediate redshifts. A cosmological constant with $w_\Lambda=-1$ is clearly ruled out  
{at more than $3\sigma$ at redshifts $z\lesssim0.24$}. The statistical preference for DE over $\Lambda$CDM is very strong, signaling a significant tension with the standard model.

Given this context, it is particularly interesting to examine how EDM performs against this dataset. As before, we begin by analyzing the best-fit behavior. The second row of Fig.~\ref{fig:best-fit} shows that, even in this case, the expansion history of the Universe in EDM closely mimics that of evolving DE. The quality of the fit is comparable, except for the EDM 2p model, which is not as flexible as the 3p and 4p models to fully capture the DES-Y5 SNe data, resulting in a slightly poorer fit.

The posterior  for the EDM EOS parameter $w_{\edm,\,0}$ { in Fig.~\ref{fig:constraints_DR2_D5}} exhibits a clear peak near the best-fit value, though this value differs slightly among the 2p, 3p, and 4p models. However, the statistical significance of this peak is only $\sim 2\sigma$ in the 2p and 3p models. In contrast, {for the 4p model we have $w_{\rm EDM,\,0}=-0.61^{+0.13}_{-0.27}$ at 95 \%CL.} Assuming a Gaussian posterior distribution for $w_{\edm,\,0}$, the 4p model shows a $\sim 3.93\sigma$ deviation of the late-time EDM EOS from that of standard CDM. While this is only a rough estimate -- the posterior distribution for this parameter is highly non-Gaussian, the more-than-3$\sigma$ detection of $w_{\edm,\,0}\neq0$ is confirmed by the predictive posterior distribution for $w_\edm(z)$ in Fig.~\ref{fig:EOS}. This improvement is tied to the smoother transition in the 4p best-fit model, where the freedom to vary the smoothing parameter $\sigma_{\edm}$ tightens the posterior on $w_{\edm,\,0}$. As in the previous analysis, the best-fit values of $\tilde{\Omega}_m$ remain high -- reaching up to $\tilde{\Omega}_m = {0.94}$ in the 4p model. Only the 2p model yields a best-fit value of $\tilde{\Omega}_m = {0.48} < 0.5$, implying that matter-DE equality would occur only at very low redshifts. The detection of an EDM component is also reflected in the tight constraints on the derived parameter $\Omega_\edm$.

Overall, while all extended models are favored over $\Lambda$CDM by this dataset, a clear hierarchy emerges. DE remains the most favored model, while EDM 2p is the least. Although EDM 3p and 4p are ``strongly" favored over $\Lambda$CDM (according to the Jeffreys' scale), they are not ``very strongly" favored, as is the case for DE. Within the uncertainty on model evidences, DE and the more flexible EDM models (3p and 4p) remain statistically consistent. This again highlights that the statistical preference for DE is driven by the higher significance in the detection of its parameters compared to those of EDM.

As for the standard cosmological parameters, both DE and EDM models yield posterior distributions for $H_0$, $\omega_b$, and derived quantities that are broadly consistent. However, $H_0$ in these models is slightly more in tension with the SH0ES result than in $\Lambda$CDM.

{Finally, we note that our constraints on the EDM 2p model are mostly consistent with those reported in Ref.~\cite{Giani:2025hhs}, which fits the DESI + DES-Y5 data using a model with the same parameterization as our EDM 2p model --- their parameter $\epsilon$ corresponds to our $w_{\edm,\,0}$ via $\epsilon = 3w_{\edm,\,0}$. (Also see footnote \ref{footnote_Giani} for comments on their model.)}
However, there are discrepancies in the inferred values of $H_0$ and $\Omega_m$, even in the $\Lambda$CDM and DE models. For the latter two models, our results align well with those of the DESI DR2 collaboration~\cite{DESI:2025zgx}, as previously discussed. The origin of the discrepancy with Ref.~\cite{Giani:2025hhs} could lie in their use of a different compressed CMB likelihood, which is based on a Gaussian prior on the 'shift parameters' $R$ and $l_A$ along with $\omega_b$~\cite{Wang:2007mza} rather than the one we adopt here (see Sec.~\ref{Sec:methodology}), or in the fact that Ref.~\cite{Giani:2025hhs} does not seem to add neutrinos to the analysis.
 
\paragraph{DR2 + CMB + Pantheon+.} We present the constraints from the analysis of this dataset in Fig.~\ref{fig:constraints_DR2_P} and summarize them in Table~\ref{tab:constraints_P+}. The best-fit predictions for the extended models are compared to the $\Lambda$CDM best fit in third row of Fig.~\ref{fig:best-fit}. 

Among the three SNe datasets considered in this work, Pantheon+ provides the weakest support for dynamical DE. In fact, the Bayes factor ``substantially disfavors" DE relative to $\Lambda$CDM. While there are still mild indications of dynamical behavior, Fig~\ref{fig:EOS} shows that the DE EOS remains consistent with a cosmological constant at the $3\sigma$ level. Moreover, the values of $\Delta\chi^2$ for the BAO and CMB components are smaller than in the DR2+CMB-only analysis (compare Table~\ref{tab:constraints_P+} with Table~\ref{tab:constraints_DR2}), suggesting that accommodating the SNe data comes at the cost of a slightly degraded fit to the BAO and CMB.

Interestingly, for this dataset, the EDM 3p and 4p models are only ``barely disfavored" and actually provide a significantly better overall fit to the full dataset compared to the DE model. This is a noteworthy feature of EDM: despite its qualitative similarity to DE in terms of expansion history, quantitatively the models can lead to distinct phenomenological predictions that are already testable with current data. This opens up the possibility of observationally distinguishing between EDM and DE, even in datasets that individually appear less supportive of dynamical DE.

\paragraph{DR2 + CMB + Union3.} We present the constraints from the analysis of this dataset in Fig.~\ref{fig:constraints_DR2_U} and summarize them in Table~\ref{tab:constraints_U3}. The best-fit predictions for the extended models are compared to the $\Lambda$CDM best fit in the fourth row of Fig.~\ref{fig:best-fit}. 

This dataset is also known to provide evidence for dynamical DE when combined with DESI DR2 BAO measurements, though the support is substantial rather than strong, as in the DES-Y5 case. Here too, the EDM models yield fits comparable to those of DE, but no statistically significant detection of EDM parameters is found.

Finally, we highlight an important caveat regarding model selection and its impact on our conclusions about EDM. As our estimator, we used the Bayes factor, which is known to depend on the choice of priors {if the priors are modified in regions of non-negligible likelihood}. This is particularly relevant for EDM due to our prior on the EOS parameter $w_{\edm,\,0}$. As shown in Section~\ref{Sec:EDM models} (see Fig.~\ref{fig:3-para-models}), both the DESI anomaly and SNe data are best fit by negative values of $w_{\edm,\,0}$. While positive values are not excluded — posterior distributions show extended, though insignificant, tails in that region — they yield poorer fits and contribute to wasted prior volume, thereby penalizing the Bayes factor.

We tested that restricting the prior to $w_{\edm,\,0} \in [-1,\,0]$ improves the Bayes factor, modifying $\ln B$ by a value ranging from $\sim -0.3$ (without SNe) to as much as $-1.0$ (with DES-Y5), potentially elevating the ranking of EDM on the Jeffreys' scale. Nevertheless, given the limited prior knowledge of EDM, we adopt a broader prior as a more conservative choice.

\section{Conclusions and discussions}
\label{Sec:Conclusions discussions}

In this paper, we have formulated and analyzed several of the simplest models of EDM, and compared them to CMB, BAO, and supernova data, to quantitatively assess their ability, in comparison to the $\Lambda$CDM and DE model, in addressing the tension recently reported by DESI DR2 between observational results on late-time cosmology and the standard model prediction.
{To answer the question of which models are better given the data, we computed Bayes factors, which automatically take into account the full parameter space explored (look-elsewhere effect) and the model complexity (penalty for additional parameters). We minimize the prior dependence by conservatively choosing the prior ranges.}\footnote{{Complementarily, we also provided the values of AIC as an example of information criteria, which answers a different question: given the model, how well it is expected to explain new data drawn from the same underlying process. The prior dependence in information criteria disappears at the expense of relying on certain asymptotic assumptions about the model distribution that may not strictly hold for the models under consideration.}}

Among the three EDM models considered in this paper, the 3-parameter model in Sec.~\ref{Sec:EDM models} --- described by the abundance of EDM, its present day EOS, and the redshift of the transition in EOS -- performs the best across different datasets. The width in the EOS step-function --- connected to the duration of the transition --- in the 4-parameter model is poorly constrained and does not add much difference at the current stage according to the Bayes factors. The 2-parameter model does not perform as well in all four combinations of datasets considered in this paper {according to the Bayes factors (but appears to be somewhat better according to AIC)}. 

Regarding the evidences against the $\Lambda$CDM model, they are only present when the Union3 or DES-Y5 supernova data is included (stronger with DES-Y5), in either the DE or EDM models. On the other hand, when including instead the Pantheon+ supernova data or without supernova data, there is no or negative evidence against the $\Lambda$CDM model. This is well consistent with the fact, pointed out in~\cite{Efstathiou:2024xcq}, that conclusions about the preference of extended models over $\Lambda$CDM strongly depend on the SNe dataset used in the analysis.

Between the DE model and the 3-parameter EDM model, the DE model is slightly better in the no-supernova case, Union3 or DES-Y5 case, while the EDM model is slightly better in the Pantheon+ case. {Therefore, the EDM model offers a comparable alternative to the DE model as an explanation for the anomaly recently claimed in DESI DR2, but without violating the null energy condition.}

The values of $H_0$ predicted by the EDM model are very similar to those from the $\Lambda$CDM or the DE model when any supernova dataset is included. Without the supernova data, the central value of $H_0$ in the DE model is shifted to a much lower value, while that of the EDM model remains similar to the $\Lambda$CDM model. So, in all cases, the tension with some of the local measurements still remains.

EDM models are also remarkable examples in which the energy budget of the present-day universe could be dramatically different from the standard model, while still remaining consistent with observations. In addition, EDM models offer an explanation for a coincidence problem in the DE model, in which the phantom-crossing and dark energy-dark matter equality occur around the same time.

There are many issues remaining to be investigated. For example:

$\bullet$ {\em Full model construction and data analysis.}
Similarly as demonstrated in \cite{Chen:2025wwn}, a simple microscopic construction of such an EDM model can be given by a canonically normalized scalar field with a potential,
\begin{align}
    V(\phi) = 
    \begin{cases}
        \frac{1}{2} m^2 \phi^2 \left( 1+ \left| \frac{\phi_1}{\phi} \right|^{\frac{4w_{\edm,\,0}}{1-w_{\edm,\,0}}} \right)^{-1} ~, & \text{for } 0<w_{\edm,\,0}<1 ~, \\    
        \frac{1}{2} m^2 \phi^2 \left( 1+ \left| \frac{\phi_1}{\phi} \right|^{\frac{-4w_{\edm,\,0}}{1-w_{\edm,\,0}}} \right) ~, & \text{for } -1<w_{\edm,\,0}<0 ~,
    \end{cases}
    \label{Eq:EDM potential}
\end{align}
where $\phi_1$ is a positive constant and $w_{\edm,0}$ is a constant. The coherent oscillation of this scalar field behaves as a species of EDM. The EOS of this EDM is density-dependent. When the oscillation amplitude is much larger than $\phi_1$ and hence its density is relatively high, the overall shape of the potential is $\propto m^2\phi^2$ and the EOS of the EDM is zero. When its density becomes lower and consequently the oscillation amplitude smaller than $\phi_1$, the shape deformation at the bottom of the potential, previously negligible, becomes important, and the EOS of this EDM approaches $w_{\edm,\,0}$. In fact, there are other types of deformation which can produce the same EOS or the same average EOS.
Envisioning a complicated dark matter potential landscape, such small deformations seem pretty natural and could exist generally.
We leave the data analysis of such full models to future work.  In this regard, we also remark that such a microscopic construction could also give a prediction for the (time-dependent) speed of sound of the EDM perturbations.

$\bullet$ {\em Other EDM models.}
Here we have only studied some of the simplest models of EDM. Given its rich possibility, it would be interesting to investigate more complicated models, especially in exploration of its connection with the Hubble tension problem or future data, because the EDM models provide a great amount of flexibility in modeling late-time cosmology while remaining microscopically easy to construct.

$\bullet$ {\em Impact on structure growth.}
{In this paper, we have only studied the effect of EDM on the background expansion. Non-standard dark matter models also change the results of perturbation theory, which can be probed by a variety of observables, such as CMB lensing and distribution of galaxies, see e.g.~\cite{Hu:1998kj,Sherwin:2011gv,Kopp:2016mhm} for analyses on generalized dark matter models or observational constraints. Because, in EDM models, non-zero EOS only emerges at late-times, most of previous constraints in the literature on the value of EOS do not apply, except for those using model-independent approaches~\cite{Kopp:2018zxp,Abedin:2025dis}, which, at late-times, broadly allow values of EOS of similar orders of magnitude as required in this work. It is important to perform dedicated analyses and constraints for the EDM models including these observables, and we will return to this in a future publication.}

$\bullet$ {\em Distinguishing models.}
Although we have shown that the EDM and DE models are currently statistically comparable in explaining the DESI anomaly, their precise predictions differ. Anticipating future higher quality data and/or more thorough data analyses from different perspectives, it is important to further work out the differences in a variety of cosmological observables such as BAO and the matter power spectrum {from both background and perturbation theory,} both within the class of EDM models and comparing to other classes of models \cite{Khoury:2025txd, Luu:2025fgw,Urena-Lopez:2025rad, Lin:2025gne, Chen:2025mlf, An:2025vfz, Cai:2025mas,Bedroya:2025fwh, Mishra:2025goj}.

$\bullet$ {\em Local observational studies of EDM.}
For models such as \eqref{Eq:EDM potential}, the properties of EDM depend on its density. As a result, regions of high dark matter density, traditionally optimal for directly observing and studying dark matter, such as galaxies and clusters, are no longer relevant for constraining the properties of EDM that influence late-time cosmological evolution. Instead, low-density environments become observationally relevant. These include cosmic voids, where the dark matter density is comparable to, or even much lower than, the cosmic average. Such regions may provide useful constraints on EDM properties relevant to present-day cosmology, and offer insight into its behavior over cosmological timescales in the future.

\medskip
\section*{Acknowledgments}

We would like to thank Antony Lewis for a comment about the likelihood in Eq.~\eqref{eq:prior_thetas}. AL was supported in part by the Black Hole Initiative, which is funded by GBMF anf JTF.

\bibliographystyle{JHEP}
\bibliography{Ref}

\end{document}